\documentclass[aps,prf,superscriptaddress,longbibliography,floatfix]{revtex4-2}
\usepackage{amsmath,amssymb,graphicx,float,epstopdf}
\usepackage{dcolumn}
\usepackage{bm}
\usepackage{color}
\graphicspath{{figs/}}
\usepackage{setspace}

\usepackage{tikz}
\usetikzlibrary{arrows.meta,bending,positioning,intersections}
\usepackage{pgfplots}
\usepackage[normalem]{ulem}

\usepackage[all]{xy}

\begin{document}

\title{Edge contact lubrication} 

\author{Andrew Wilkinson}
\email[]{andrew.wilkinson@open.ac.uk,mmorgan@seattle.edu,m.wilkinson@open.ac.uk}
\affiliation{School of Mathematics and Statistics, The Open University, Milton Keynes MK7 6AA, United Kingdom}
\author{Michael A. Morgan}
\affiliation{Department of Physics, Seattle University, Seattle, WA 98122, USA}
\author{Michael Wilkinson}
\affiliation{School of Mathematics and Statistics, The Open University, Milton Keynes MK7 6AA, United Kingdom}

\date{\today}

\begin{abstract}
In this paper we discuss the dynamics of a lubricating film separating two flat
surfaces. It may be possible for the gap between the surfaces to reduce to zero 
at one point on the boundary in a finite time. Typically, the maximum pressure diverges as the gap between the surfaces 
approaches zero. We discuss the form of the pressure field in the vicinity of its 
maximum. We also introduce a numerical method for accurate determination of the 
pressure across the lubrication surface, enabling us to determine the hydrodynamic 
forces. We determine the phase diagram describing the evolution of the gap, as a function 
of the initial configuration and of the line of action of the external forces. When contact 
does occur, we consider whether the subsequent motion involves sliding of the point of 
contact. We argue that, when sliding does occur, the motion does not depend upon the 
microscopic properties of the surface roughness.  
\end{abstract}

\maketitle 

\section{Introduction}
\label{sec: 1}

Lubrication theory \cite{Rey86,Mic50,Hap+83,Kim+91,Sze98}   
describes how a film of viscous fluid can resist a force which 
drives two surfaces together. The thickness of the film layer may decrease without 
the surfaces ever making contact (in the context of a continuum approximation). 
It is also possible that the surfaces may make contact in finite time. This latter case 
is the topic of this paper. We study the case 
where the edge of an object with a flat lower surface is close to making contact 
with a flat planar surface below it. The motion may, or may not, involve the surfaces 
making contact after a finite time. If there is contact, subsequent motion may involve pivoting 
about the point of contact, causing the lubricating fluid to be forced out of the gap. In other 
cases pivoting about a fixed contact point is untenable, and the contact point must move. 
Our objective is to explore the long-time ($t\to \infty$) dynamics of this system.

This topic is relatively unexplored, but we remark that \cite{Bre61,Gol+67,Gol+67a,Jef+84,Jef+84a} 
discuss the close approach of a sphere to surfaces, and other spheres, and that \cite{Caw+10,Bal+10} 
discuss a wedge-shaped object contacting a flat surface.
A special case of this problem was addressed in an earlier work \cite{Wil+23}, where we considered the motion 
of a straight-edged plate settling onto a flat surface, in two dimensions. We found that (depending upon 
initial conditions and the position of the centre of gravity) in the long-time limit the plate may 
approach a flat configuration, with a gap which decreases as $t^{-1/2}$, or alternatively the plate 
may make contact in a finite time. In cases where contact occurs in finite time, pivoting 
about a fixed contact point occurs if the centre of gravity is close to the edge which makes contact. 
In other cases, it was concluded that the point of contact must slide as the lubricating fluid is 
squeezed out of the gap. It was argued that, in the case of a straight edge, the sliding motion after 
contact depends upon the microscopic structure of the surfaces. 

However, a two-dimensional treatment is not the most relevant case for real-world applications, 
because it is unlikely that contact will be made simultaneously along a straight line edge. In this work 
we consider cases where the settling object has a curved edge or a corner, so that contact occurs at a 
point, rather than along a line. This differs significantly from the case treated in \cite{Wil+23}. 
It is argued that, in this case, the sliding motion after contact is determined by lubrication 
theory alone, without requiring any information about the microscopic structure of the surfaces.
We emphasise that our work treats the problem purely in the framework of a continuum model, in which surfaces are 
perfectly flat and rigid, and where friction between contacting surfaces is in accord with the Coulomb model. We do not 
consider processes which occur when the lubricating film has been squeezed to the length scale of surface roughness.

The central difficulty in solving problems in lubrication theory is to determine the pressure field in the 
gap between the surfaces. This satisfies a linear partial differential equation obtained by Reynolds \cite{Rey86}. 
In cases where the gap between the surfaces is not constant in space, this equation is difficult to solve analytically, except 
in two dimensions. Here we use two complementary approaches to address this difficulty. 
We introduce a novel numerical method for solving the Reynolds equation, based upon 
minimisation of a functional. We also use the observation that the pressure becomes very large 
in the region where the gap is very small. In the limiting case where the gap approaches zero we 
give an asymptotic approximation to the pressure field, using a modification of the two-dimensional
solution given in \cite{Wil+23}.

\begin{figure}
\centering
\includegraphics[width=0.65\textwidth]{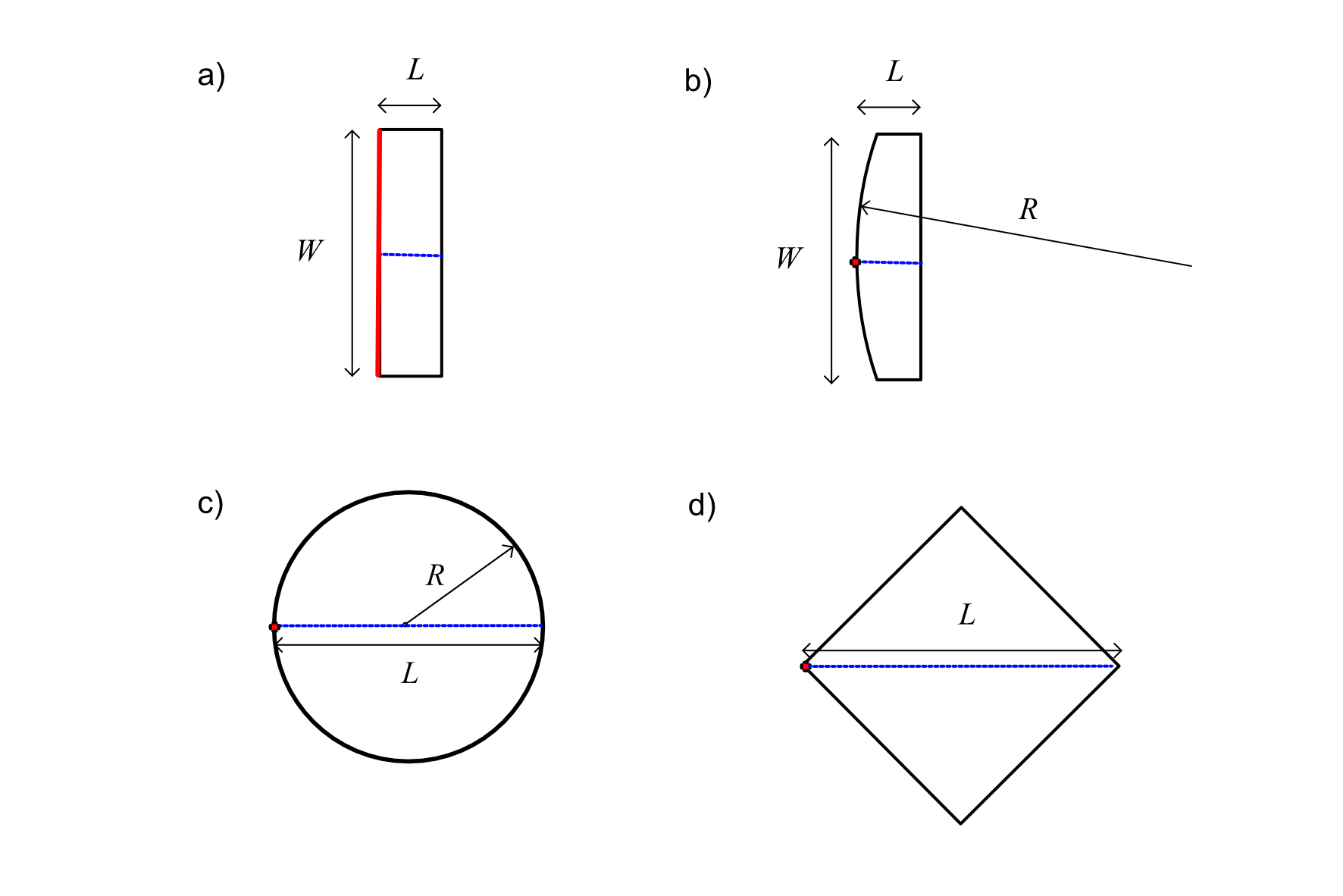}
\caption{
Illustrating the different forms for the upper, freely moving surface. Points closest to contact 
are highlighted with a red dot or line. All examples are axisymmetric: the line of symmetry
is indicated by a dotted blue line. ({\bf a}) Rectangle. ({\bf b}) Rectangle with curved boundary.
({\bf c}) Disc. ({\bf d}) Square with diagonal axisymmetry.
}
\label{fig: 1}
\end{figure}

In order to avoid some complications which arise in the most general case,
this paper will be restricted to considering cases where there is an axis of 
symmetry passing through the point at which the gap is minimal. Four {\it axisymmetric} cases,
illustrated in figure \ref{fig: 1}, will be discussed in detail. In each case we take the length of the object along its 
symmetry axis to be the characteristic length $L$. Case ({\bf a}) is a rectangular 
plate, with aspect ratio $W/L$, with a constant small gap at its left-hand edge. This is included as a means
to validate the numerical method, by showing that the results approach the two-dimensional case 
treated analytically in \cite{Wil+23} in the limit as $W/L\to \infty$.  Case ({\bf b})
is a rectangle with a curved edge, with radius $R$. This case is considered because, in the 
limit as $R\to \infty$, the pressure field can be approximated analytically, based 
upon the approach developed in \cite{Wil+23}. Such an approximation should be valid for other symmetrical 
objects with a large radius of curvature $R$ at the leading edge.
Case ({\bf c}) is a flat disc of radius $R$, which would be the easiest example 
for an experimental study. Case ({\bf d}) is a square with its axis of symmetry along 
the diagonal. This case was considered because it is the simplest example where 
contact occurs at a corner.  In all four cases the gap, as well
as the surface, is symmetric about the symmetry axis.

Section \ref{sec: 2} discusses the formulation of the problem, including the Reynolds
equation determining the pressure distribution. The pressure can diverge as the gap between
the surfaces approaches zero, and section \ref{sec: 3} discusses the divergent pressure 
which arises from a region where the gap is nearly closed at the edge.
In section \ref{sec: 4} we outline a novel method for solving the Reynolds equation numerically. 
These various elements are used in section \ref{sec: 5} to determine the resistance matrix which defines the equation
of motion, in the case where the edge is very close to making contact. In section \ref{sec: 6} these results are 
used to analyse the motion close to contact when the upper object is sinking under the action of gravity, both in general 
terms, and for the various examples illustrated in figure \ref{fig: 1}. We consider four possible 
generic motions that can occur: 

\begin{itemize}

\item The falling object may contact the surface in finite time, followed by continuing to sink by pivoting 
about the point of contact.

\item The falling object may make contact in finite time, after which the point of contact slides 
while the object continues to sink. We argue that the sliding motion is not influenced by the nature of 
the roughness of the surfaces which come into contact.

\item The object may sink without contact occurring in finite time.

\item The object may tilt so that the edge at the opposite 
end of the axis of symmetry is closest to contact. 

\end{itemize}

All of these possibilities may be realised, depending on the initial conditions and position of the centre of 
mass of the sinking object. In section \ref{sec: 7} we present phase diagrams, 
showing the long-time behaviour as a function of the initial configuration and the centre of mass position.
Section \ref{sec: 8} is a brief conclusion.

\section{Equations of motion}
\label{sec: 2}

\subsection{Fundamental equations}
\label{sec: 2.1}

\begin{figure}[h!]
\centering
\begin{tikzpicture}[scale=0.7]
\path[draw,line width=4pt](2,0)--(17,0);
\path[draw,line width=3.5pt](4.8,0.8)--(15,1.8);
\draw[-{Stealth[length=2mm, width = 2mm]}] (4.8,0) -- (4.8,0.78) node[pos=.5,left] {$Z(t)$};
\draw[-] (4.8,0.75) -- (9.5,0.75) node[pos=0.65,above,yshift=-0.11cm] {$\qquad \qquad \qquad \qquad \ \ \theta(t)$};
\draw[-{Stealth[length=2mm, width = 2mm]}] (4.8,1.1) -- (10,1.61) node[pos=.5,above] {$x$};
\draw[-{Stealth[length=2mm, width = 2mm]}] (3,-0.25) -- (4.8,-0.25) node[pos=.5,below] {$X(t)$};
\draw[-{Stealth[length=2mm, width = 2mm]}] (11,1.4) -- (11,0.5) node[pos=.5,right] {${\cal W}$};
\draw[-{Stealth[length=2mm, width = 2mm]}] (4.8,-0.5) -- (11,-0.5) node[pos=.5,below] {$sL$};
\draw[{Stealth[length=2mm, width = 2mm]}-{Stealth[length=2mm, width = 2mm]}] (4.75,1.5) 
-- (14.9,2.5) node[pos=.5,above] {$L$};
\draw[line width=1pt] (3,-0.4)--(3,0) node [pos=1.0,above]{$X=0$};
\draw[-{Stealth[length=2mm, width = 2mm]}] (2.2,0) -- (2.2,2.5) node[pos=.5,right] {$z$};
\draw[] (9,1) arc (0:1:15);
\draw[] (9,1) arc (0:-1:15);
\end{tikzpicture}
\caption{
A cross-section through the object along the line of symmetry axis which has length $L$. 
The minimal gap $Z(t)$ occurs the left-hand edge, 
which has displacement $X(t)$ from a fixed reference
point on the plane. The tilt angle is $\theta(t)$. The calculation uses Cartesian coordinates with their origin on the 
upper object, with the $x$-axis along the symmetry axis and the $y$-axis out of the paper. The weight ${\cal W}$ 
acts at a centre of mass which is displaced by a distance $sL$ from the left-hand edge.
}
\label{fig: 2}
\end{figure}
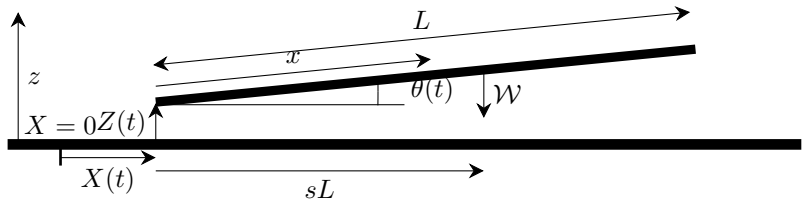

We consider two flat surfaces being forced together, resisted by a lubrication 
film. The lower surface is infinite, and the upper one is finite. 
Three generalised coordinates, $Z$, $X$, $\theta$, are required to specify the configuration, 
as illustrated schematically in figure \ref{fig: 2}: $Z$ is the minimum gap, $\theta$ is the tilt 
angle and $X$ is the horizontal displacement of the object.

The assumption that there is a Poiseuille flow in the gap implies that the 
volume flux, $\mbox{\boldmath$J$}$,  and the pressure $p$ are related by
\begin{equation}
\label{eq: 2.1}
\mbox{\boldmath$J$}=\left(\frac{\mbox{\boldmath$v$}_1+\mbox{\boldmath$v$}_2}{2}\right)h
-\frac{h^3}{12\mu}\mbox{\boldmath$\nabla$}p
\end{equation}
where $\mbox{\boldmath$v$}_1$, $\mbox{\boldmath$v$}_2$ are the velocities of the boundaries at 
(respectively) the lower and upper surfaces of the lubricating layer, and $\mu$ is the viscosity.
The continuity equation  is 
\begin{equation}
\label{eq: 2.2}
\frac{\partial h}{\partial t}+\mbox{\boldmath$\nabla$}\cdot \mbox{\boldmath$J$}=0
\ .
\end{equation}
Poiseuille flow also implies that the tangential stress on the upper surface is
\begin{equation}
\label{eq: 2.3}
\mbox{\boldmath$\sigma$}=-\mu\left(\frac{\mbox{\boldmath$v$}_2-\mbox{\boldmath$v$}_1}{h}\right)
-\frac{h}{2}\mbox{\boldmath$\nabla$} p
\ .
\end{equation}
The calculation uses Cartesian coordinates with their origin on the 
upper object, with the $x$-axis along the symmetry axis and the $y$-axis out of the paper. 
The displacement of the left-hand edge of the plate in the laboratory frame is $X(t)$, so that when the 
angle $\theta$ is small the speed of the lower surface is very well approximated by $\mbox{\boldmath$v$}_1=-\dot X{\bf e}_1$, 
while the upper surface is stationary, $\mbox{\boldmath$v$}_2={\bf 0}$. 
The gap height at position $(x,y)$ is independent of $y$:
\begin{equation}
\label{eq: 2.4}
h(x,t)=Z(t)+\theta(t)x
\end{equation}
where $\theta(t)$ is the angle between the plate and the horizontal. As usual in lubrication theory, we assume throughout 
that this gap height is smaller than any transverse dimension in the problem, and therefore that the fluid pressure $p$ is independent 
of the vertical coordinate $z$. Also, note that throughout this paper we always assume $\theta$ is sufficiently small that $\cos\theta\approx 1$ and $\sin\theta\approx\theta$.

Combining (\ref{eq: 2.1}) and (\ref{eq: 2.2}) leads to an equation for the pressure:
\begin{equation}
\label{eq: 2.5}
\frac{\partial h}{\partial t}-\frac{\dot X}{2}\frac{\partial h}{\partial x}
=\frac{1}{12\mu}\mbox{\boldmath$\nabla$}\cdot \left(h^3 \mbox{\boldmath$\nabla$}p\right)
\end{equation}
which is known as the Reynolds equation \cite{Rey86}. The Reynolds equation is to be solved with the 
boundary condition $p(x,y,t)=p_0(t)$ on the boundary of the lubrication layer, where $p_0(t)$ is the 
external pressure, which we assume is sufficient to prevent cavitation.  
Note that the Reynolds equation is a linear partial differential equation. Also, noting 
that $\partial h/\partial t=\dot Z+x\dot \theta$, we observe that 
the terms which are independent of $p(x,y,t)$ are linear combinations
of the generalised velocities, so that we may define three pressure functions according to
\begin{equation}
\label{eq: 2.6}
p(x,y,t)=P_X(x,y)\,\dot X(t)+P_\theta (x,y)\,\dot \theta(t)+P_Z(x,y)\,\dot Z(t) +p_0
\end{equation}
where the function $P_Z(x,y)\,\dot Z(t)$ is the solution of the Reynolds equation with $\dot \theta=\dot X=0$, 
and similarly for $P_\theta$ and $P_X$.

The motion of the settling object is determined by balancing external forces against 
the hydrodynamic forces due to pressure and shear stress. By symmetry, there is no
force in the $y$-direction, and no torque about the $x$ or $z$-axes. There are, therefore, 
three generalised forces, obtained by integrating pressure and stress fields over the region 
${\cal A}$ where the bodies are close to contact: 
\begin{subequations}
\label{eq: 2.7}
\begin{align}
F_z&=\int_{\cal A}\!{\rm d}\mbox{\boldmath$x$}\ p\\
F_x&=\int_{\cal A}\!{\rm d}\mbox{\boldmath$x$}\ \sigma_x \;-\;\theta F_z\\
G_y&=\int_{\cal A}\!{\rm d}\mbox{\boldmath$x$}\ xp \;.
\end{align}
\end{subequations}

The reference axis for $G_y$ is the line $x=z=0$, and it is measured anticlockwise 
from the perspective of figure \ref{fig: 2}.

Noting that the pressure field is linear in the generalised velocities, the equation (\ref{eq: 2.7}) can be 
used to express the vector of generalised forces, $\mbox{\boldmath$F$}=(F_x,F_z,G_y)$ in terms of the 
generalised velocity vector $\dot{\mbox{\boldmath$X$}}=(\dot X,\dot Z,\dot \theta)$: we write
\begin{equation}
\label{eq: 2.8}
\mbox{\boldmath$F$}={\bf R}(\mbox{\boldmath$X$})\dot{\mbox{\boldmath$X$}}
\end{equation}
where the \emph{resistance matrix} ${\bf R}$ is a $3\times 3$ matrix with elements which are 
functions of the generalised coordinate vector $\mbox{\boldmath$X$}=(X,Z,\theta)$.

\section{Asymptotic forms for pressures}
\label{sec: 3}

In cases where an edge is very close to making contact, 
the pressure will typically become extremely large, diverging as the edges come into contact. This section discusses
the form of the pressure distribution in the limit as the edge separation approaches zero. 

We discuss the case illustrated in figure \ref{fig: 1}({\bf b}), in the limit as $R\to \infty$. 
Consider the pressure at a distance $x$ along a line parallel to the axis of symmetry, 
with a displacement $y$ from this axis. 
In order to simplify the boundary condition, in this section we use a coordinate system 
with $x=0$ at the leading edge of the object (so that $x=L-f(y)$ is its trailing edge). 
For the example defined in figure \ref{fig: 1}({\bf b}), the edge profile is
\begin{equation}
\label{eq: 3.1}
f(y)=\frac{y^2}{2R}
\ \ ,\ \ \ -\frac{W}{2} \le y\le \frac{W}{2} 
\ .
\end{equation}
Because $W\gg L$, and the profile of the leading edge of the plate varies very 
slowly as a function of $y$, we neglect fluid flow in the $y$-direction, so that the fluid flow will be 
treated as one-dimensional. The $y$ coordinate appears as a parameter because the gap at the leading edge
depends upon $y$, according to ${\cal Z}\equiv Z+\theta f(y)$. As a simplification we will ignore the dependence of 
the plate length upon the value of $y$, taking the trailing edge to be at a constant  $x=L$ rather than the 
variable $x=L-f(y)$, since this will have negligible effect on the results.

\subsection{Approximating the pressure field}
\label{sec: 3.1}

Treating the $y$-coordinate as a parameter, equations (\ref{eq: 2.1}), (\ref{eq: 2.2}) and (\ref{eq: 2.4}) 
are then replaced by a one-dimensional system:
\begin{eqnarray}
\label{eq: 3.1.1}
&&J=-\frac{\dot Xh}{2}-\frac{h^3}{12\mu}\frac{\partial p}{\partial x}
\nonumber \\
&&\frac{\partial h}{\partial t}+\frac{\partial J}{\partial x}=0
\nonumber \\
&&h(x,y,t)={\cal Z}(y,t)+\theta(t)x \ ,\ \ \ {\cal Z}(y,t)=Z(t)+\theta(t)\, f(y)
\ .
\end{eqnarray}
Using the expression for $h(x,y,t)$ in the continuity equation and integrating once we obtain
$J=J_0-x\dot {\cal Z}-\frac{1}{2}x^2\dot \theta$, where $J_0$ is a constant of integration.
Hence, using the first equation of (\ref{eq: 3.1.1}), the pressure gradient is
\begin{equation}
\label{eq: 3.1.2}
\frac{\partial p}{\partial x}=12\mu\left[\frac{x}{h^3}\dot {\cal Z}+\frac{x^2}{2h^3}\dot \theta-\frac{1}{2h^2}\dot X-\frac{1}{h^3}J_0\right]
\end{equation}
where $J_0(y)$ depends parametrically upon $y$. It will be useful to define
\begin{equation}
\label{eq: 3.1.3}
I^n_m(x,y)\equiv \int_0^x {\rm d}x'\ \frac{x'^n}{[h(x',y)]^m}
\ .
\end{equation}
Integrating (\ref{eq: 3.1.2}) and imposing the boundary condition $p(0,y)=p(L,y)=p_0$ leads to an 
expression for the constant of integration $J_0(y)$ and hence 
\begin{eqnarray}
\label{eq: 3.1.5}
\frac{p(x,y)-p_0}{12\mu}I_3^0(L,y)&=&\dot {\cal Z}\left[I^1_3(x,y)I^0_3(L,y)-I^0_3(x,y)I^1_3(L,y)\right]
\nonumber \\
&+&\frac{\dot \theta}{2}\left[I_3^2(x,y)I^0_3(L,y)-I^0_3(x,y)I^2_3(L,y)\right]
\nonumber \\
&-&\frac{\dot X}{2}\left[I^0_2(x,y)I^0_3(L,y)-I^0_3(x,y)I^0_2(L,y)\right]
\ .
\end{eqnarray}
Consider the evaluation of the integrals $I^n_m$ appearing in (\ref{eq: 3.1.5}). Introducing the dimensionless variable
\begin{equation}
\label{eq: 3.1.6}
\xi=\frac{\theta x}{{\cal Z}}
\end{equation}
$I^0_3(x,y)$ can be evaluated as follows:
\begin{equation}
\label{eq:  3.1.7}
I^0_3(x,y)=\int_0^x \frac{{\rm d}x'}{({\cal Z}+x'\theta)^3}=\frac{1}{{\cal Z}^2\theta}\int_1^{1+\xi}\frac{{\rm d}v}{v^3}
=\frac{1}{{\cal Z}^2\theta}\frac{\xi(2+\xi)}{2(1+\xi)^2}
\equiv \frac{1}{{\cal Z}^2\theta} g(\xi)
\ .
\end{equation}
Similarly, the others are
\begin{eqnarray}
\label{eq: 3.1.8}
I^1_3(x,y)&=&\frac{1}{{\cal Z}\theta^2}\frac{\xi^2}{2(1+\xi)^2}\equiv \frac{1}{{\cal Z}\theta^2}f_Z(\xi)
\nonumber \\
I^2_3(x,y)&=&\frac{1}{\theta^3}\left[\ln(1+\xi)-\frac{(3\xi+2)\xi}{2(1+\xi)^2}\right]
\equiv \frac{1}{\theta^3}f_\theta(\xi)
\nonumber \\
I^0_2(x,y)&=&\frac{1}{{\cal Z}\theta}\frac{\xi}{1+\xi}
\equiv \frac{1}{{\cal Z}\theta} f_X(\xi)
\ .
\end{eqnarray}
Hence, in the limit $\eta\equiv L\theta/{\cal Z}\gg 1$,
\begin{eqnarray}
\label{eq: 3.1.9}
P_Z(x,y)&=&\frac{12\mu}{{\cal Z}\theta^2}\left[f_Z(\xi)-\frac{g(\xi)}{g(\eta)}f_Z(\eta)\right]
\nonumber \\
&\sim&-\frac{12\mu}{{\cal Z}\theta^2}\frac{\xi}{(1+\xi)^2} \left(1-\frac{x}{L}\right)
\nonumber \\
P_\theta(x,y)&=&\frac{6\mu}{\theta^3}\left[f_\theta(\xi)-\frac{g(\xi)}{g(\eta)}f_\theta(\eta)\right]
\nonumber \\
&\sim&\frac{6\mu}{\theta^3}\left[\ln[1+\xi]-\frac{\xi(2+\xi)}{(1+\xi)^2}\ln[1+\eta]+\frac{2\xi}{(1+\xi)^2} 
\frac{Z}{{\cal Z}}\left(1-\frac{x}{L}\right)\right]
\nonumber \\
P_X(x,y)&=&-\frac{6\mu}{{\cal Z}\theta}\left[f_X(\xi)-\frac{g(\xi)}{g(\eta)}f_X(\eta)\right]
\nonumber \\
&\sim&\frac{6\mu}{{\cal Z}\theta}\frac{\xi}{(1+\xi)^2} \left(1-\frac{x}{L}\right)
\ .
\end{eqnarray}

The pressure may be written in the form 
\begin{equation}
\label{eq: 3.2.3}
p(x,y)=6\mu \left[\frac{\dot X}{\theta} - \frac{2\dot Z}{\theta^2}+\frac{2Z\dot \theta}{\theta^3}\right]Q_1(x,y)
+\frac{6\mu \dot \theta}{\theta^3}Q_2(x,y)+\Delta p(x,y)
\end{equation}
where the fields $Q_1(x,y)$ and $Q_2(x,y)$ are  
\begin{eqnarray}
\label{eq: 3.2.1}
Q_1(x,y)&\equiv&\frac{1}{{\cal Z}}\frac{\xi}{(1+\xi)^2} \left(1-\frac{x}{L}\right)
=\frac{\theta x}{\left[Z+\left(x+f(y)\right)\theta\right]^2} \left(1-\frac{x}{L}\right)
\nonumber \\
Q_2(x,y)&\equiv&\ln(1+\xi)-\frac{\xi(2+\xi)}{(1+\xi)^2}\ln(1+\eta)
\nonumber \\
&=&\ln\left[\frac{Z+\left(x+f(y)\right)\theta}{Z+f(y)\theta}\right]
-\frac{x\theta\left[2Z+\left(x+2 f(y)\right)\theta\right]}{\left[Z+\left(x+f(y)\right)\theta\right]^2}
\ln\left[\frac{Z+\left(L+f(y)\right)\theta}{Z+f(y)\theta}\right]
\ .
\end{eqnarray}
Here $\Delta p(x,y)$ is a correction which compensates for the calculation 
ignoring flow in the $y$-direction, and could be determined numerically if required. 
The functions $Q_1(x,y)$ and $Q_2(x,y)$ are divergent as $Z\to 0$, 
whereas $\Delta p(x,y)$ remains bounded as $Z\to 0$. 

In the ${\cal Z}\ll L$ limit which we are considering, the field $Q_1(x,y)$ 
has its maximum value at $x=Z/\theta+f(y)$, where this maximum
is approximately equal to $1/4{\cal Z}$. The asymptotic behaviour of the field $Q_2(x,y)$ 
is more complicated. In the limit as $\eta\to \infty$, its minimum, located at $\xi\approx\sqrt{2\, \ln \eta}-1$, 
is to leading order, $-\ln \eta$. 

\section{Numerical determination of the pressure field and forces}
\label{sec: 4}

\subsection{Numerical method}
\label{sec: 4.1}

In order to determine numerical solutions to equations (\ref{eq: 2.5}), consider the functional
\begin{equation}
\label{eq: 4.1}
S[p]=\int_{\cal A}{\rm d}\mbox{\boldmath$x$}\;\left\{\frac{1}{24\mu}[h(x,y)]^3|\mbox{\boldmath$\nabla$}p|^2+f(x,y)p \right\} 
\ .
\end{equation}
The Euler-Lagrange equation for this functional is the Reynolds equation, (\ref{eq: 2.5}), with 
\begin{equation}
\label{eq: 4.2}
f(x,y)=\frac{\partial h}{\partial t}-\frac{\dot X}{2}\frac{\partial h}{\partial x}=\dot Z+x \dot \theta -\frac{\theta}{2}\dot X
\ .
\end{equation}
The functional (\ref{eq: 4.1}) is quadratic and positive definite, so we could determine the function $p(x,y)$ by minimising 
$S[p]$ subject to the boundary condition that $p(x,y)=p_0=0$ on the boundary. Note that in the work presented here, where we 
are assuming an axis of symmetry, $h$ and $f$ are only functions of $x$, but the numerical method described below would 
also work for the more general case of 2d flow.

To solve the equation numerically, we make a Delaunay triangularisation of the region ${\cal A}$, with ${\cal N}\gg 1$ vertices
in the interior, and ${\cal M}$ vertices on the boundary. We used a triangularisation with smaller triangles 
close to the left-hand edge, where the pressure is expected to vary more rapidly. At each vertex $k$ the pressure 
field is represented by a number $p_k$.
The functional is represented by the sum of contributions from each triangle, each of which are quadratic in the values 
of $p_k$ at the three vertices, here denoted by $p_{\rm a}$,  $p_{\rm b}$, $p_{\rm c}$:
\begin{equation}
\label{eq: 4.3}
S[p]\sim \sum_{\rm triangles} \Delta S_{\rm triangle}
\ ,\ \ \ 
\Delta S_{\rm triangle}=\Delta A\left[\sum_v\sum_{v'} T_{v,v'}p_vp_{v'}+\tfrac{1}{3}\bar f (p_{\rm a}+p_{\rm b}+p_{\rm c})\right] 
\end{equation}
where $v,v'\in\{{\rm a,b,c}\}$, $\bar f$ is the mean value of $f(x,y)$ at the three vertices, and where the coefficients $T_{v,v'}$ 
depend on the geometry of the triangle, and $\Delta A$ is the area of the elementary triangle. 
The method will be discussed in detail in a later work, which will address the solution 
of the Reynolds equation in some generality. 

Adding the contributions from all of the triangles, the functional is estimated by
\begin{equation}
\label{eq: 4.4}
S[p]=\sum_{j,k=1}^{\cal N} C_{jk} p_j p_k+\sum_{k=1}^{\cal N}D_k p_k
\end{equation}
where the sums run over all of the ${\cal N}$ internal vertices and where the elements $C_{jk}$ and $D_j$ are obtained from the
sum over triangles in equation (\ref{eq: 4.3}). The values $p_k$ which minimise (\ref{eq: 4.4}) are determined 
by solving the system of ${\cal N}$ linear equations:
\begin{equation}
\label{eq: 4.5}
\sum_{k=1}^{\cal N} C_{jk}p_k=-\tfrac{1}{2}D_j
\ .
\end{equation}
Having thus found the $p_k$, we approximate the integrals in (\ref{eq: 2.7}) by
summing over the Delaunay triangles. 
 
\subsection{Examples}
\label{sec: 4.2}

Here we present some examples of the numerically determined pressure fields, in order to 
demonstrate the utility of the method and to verify the asymptotic approximations 
of section \ref{sec: 3.1}. Note that from this point in the paper we will restrict the definition of $\eta$ to the symmetry axis, 
namely, from now on $\eta\equiv L\theta/Z$.

\begin{figure}[h]
\includegraphics[width=1\textwidth]{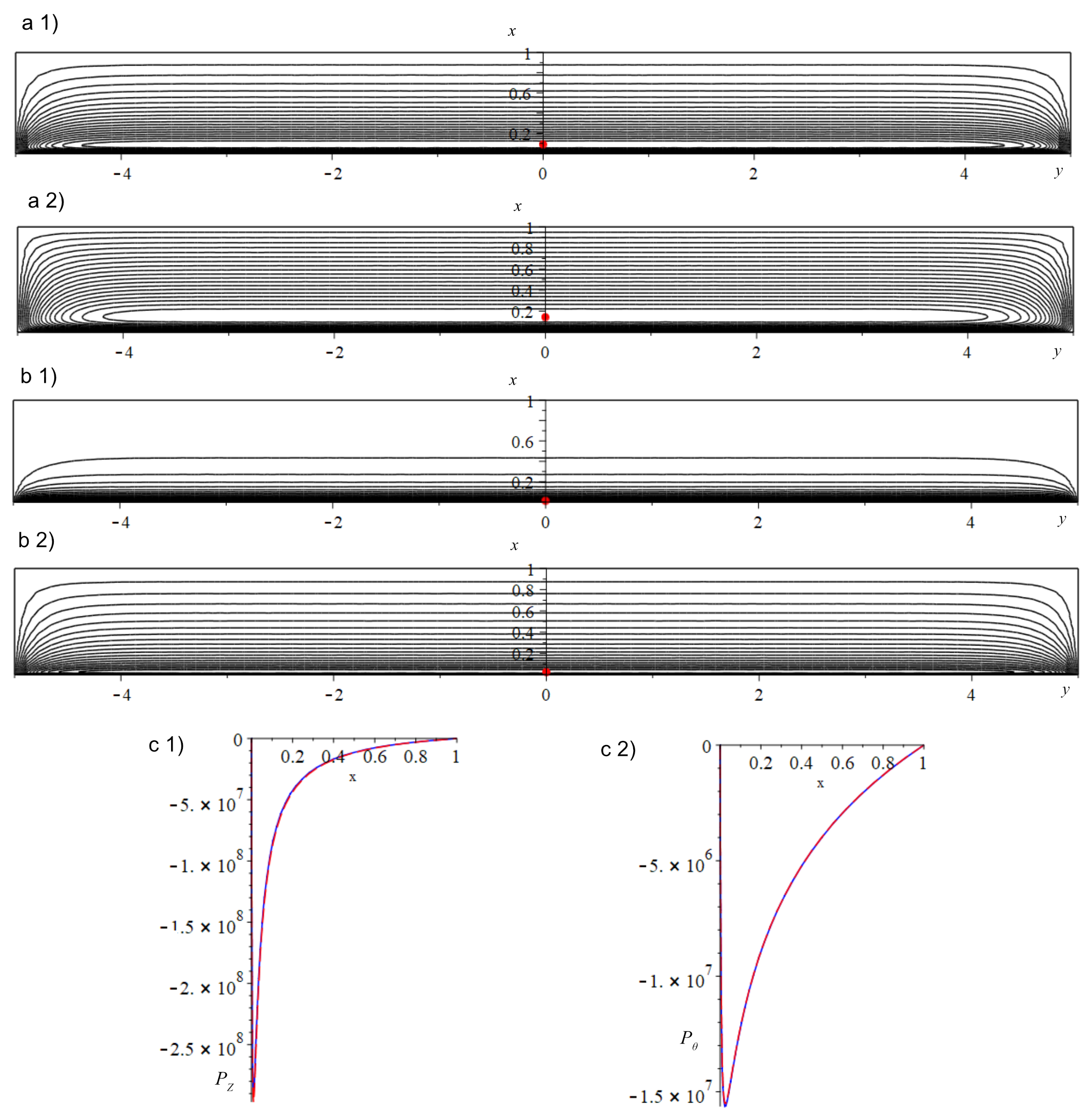}

\caption{
Pressure field plots for the rectangular 
plate illustrated in figure \ref{fig: 1}({\bf a}), with aspect ratio $W/L=10$. 
(a1) and (a2) show contour plots of $P_Z$ and $P_\theta$ respectively for 
$\theta=0.01$, $Z=0.001$. (b1) and (b2) show contour plots of $P_Z$ and $P_\theta$ 
respectively for $\theta=0.01$, $Z=0.0001$. Red dots indicate pressure maxima. 
(c1) and (c2) show pressure along the $x$-axis for $P_Z$ and $P_\theta$ respectively in 
the $Z=0.0001$ case. Here our numeric results (in blue) are compared with the predictions 
of equation (\ref{eq: 3.1.9}) (in red). 
}
\label{fig: 3}
\end{figure}

We determined the pressure fields $P_Z$ and $P_\theta$ for the rectangular 
plate of figure \ref{fig: 1}({\bf a}), with aspect ratio $W/L=10$, 
for parameter values $L=1$, $\theta=0.01$, and $Z=0.001$, $Z=0.0001$,
corresponding to $\eta=10$ and $\eta=100$, respectively.
We plot the pressure contours and the pressure along the $x$-axis for both fields $P_Z$, $P_\theta$ 
in figure \ref{fig: 3}. We compare the pressure along the $x$-axis, for the $\eta=100$ case, 
to the asymptotic forms given in equations (\ref{eq: 3.2.3}) and (\ref{eq: 3.2.1}), with $\Delta p=0$. 
In this case we find excellent agreement, because the pressure gradient in the $y$-direction is negligible.

\begin{figure}[h]
\includegraphics[width=0.8\textwidth]{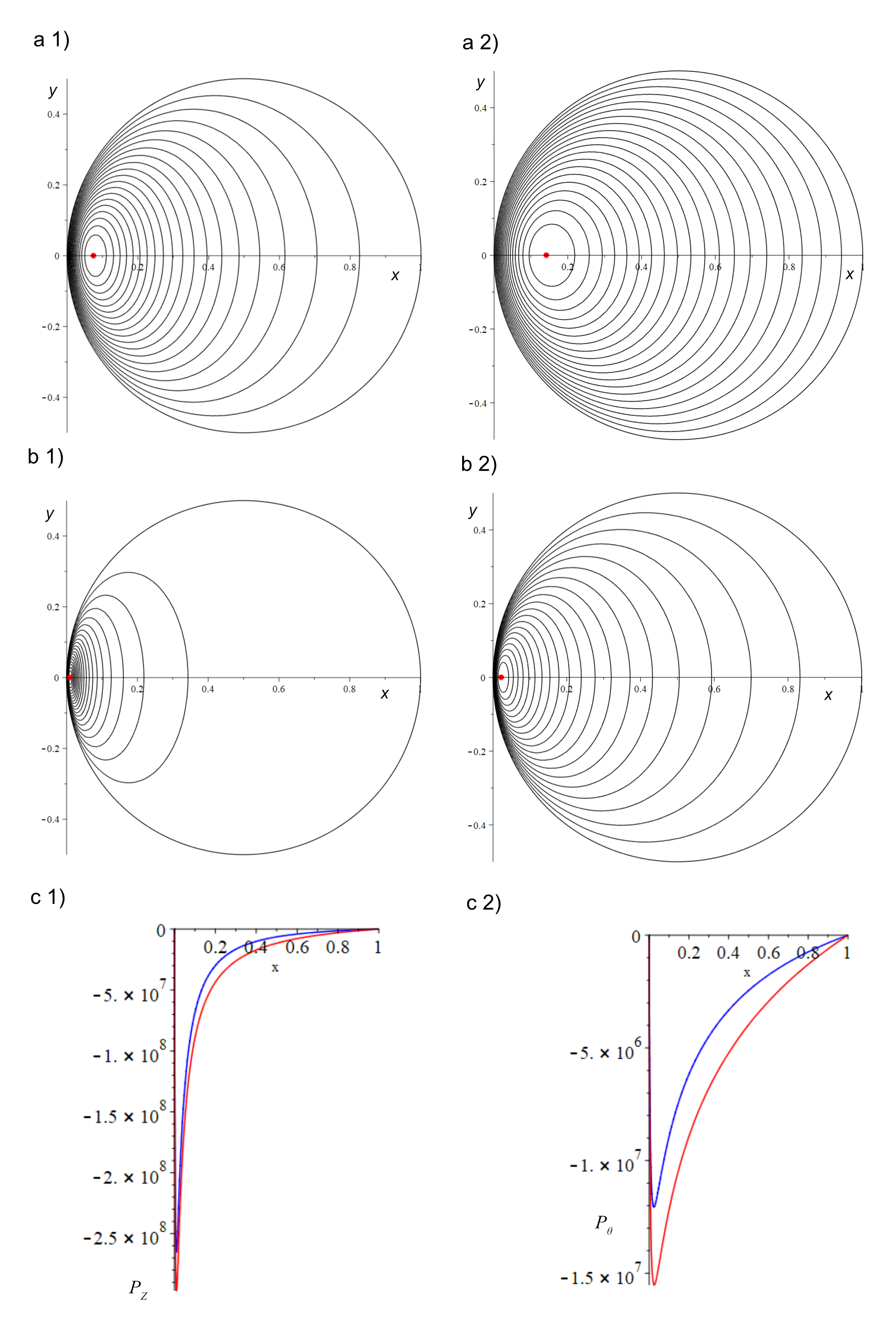}

\caption{
Pressure field plots for the disc illustrated in figure \ref{fig: 1}({\bf c}), with $R=1/2$. 
(a1) and (a2) show contour plots of $P_Z$ and $P_\theta$ respectively for 
$\theta=0.01$, $Z=0.001$. (b1) and (b2) show contour plots of $P_Z$ and $P_\theta$ 
respectively for $\theta=0.01$, $Z=0.0001$. Red dots indicate pressure maxima. 
(c1) and (c2) show pressure along the $x$-axis for $P_Z$ and $P_\theta$ respectively 
in the $Z=0.0001$ case. Here our numeric results (in blue) are compared with the predictions 
of equation (\ref{eq: 3.1.9}) (in red).
}
\label{fig: 4}
\end{figure}

\begin{figure}[h]
\includegraphics[width=0.8\textwidth]{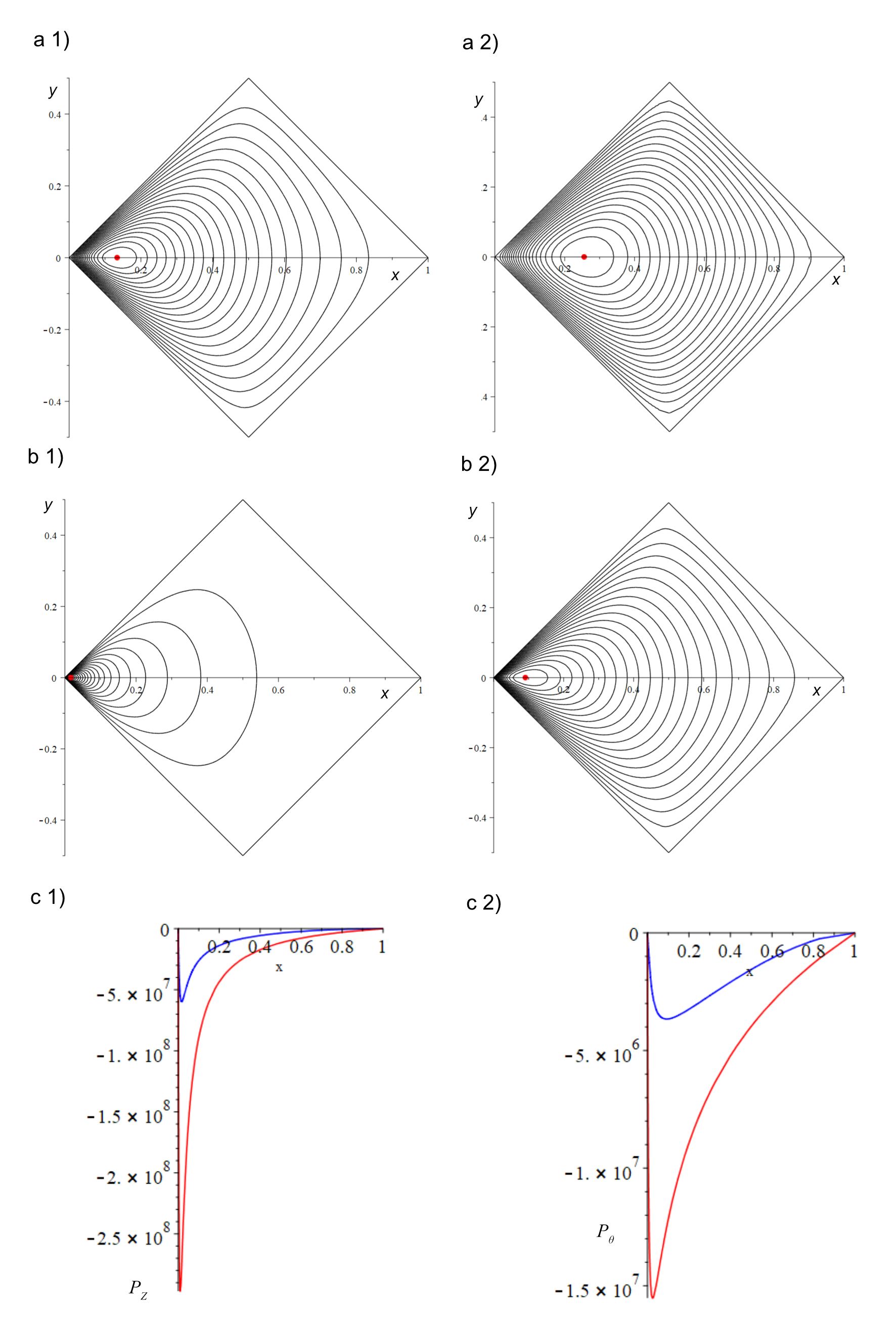}
\caption{Pressure fields $P_Z$, $P_\theta$  for the diagonal square of figure \ref{fig: 1}({\bf d}), with $L=1$: 
same format as figure \ref{fig: 4}.
}
\label{fig: 5}
\end{figure}

Figure \ref{fig: 4} shows contours of the fields $P_Z$ and $P_\theta$ for the disc of figure \ref{fig: 1}({\bf c}), with 
$R=1/2$, $\theta=0.01$ and $Z=0.001$, $Z=0.0001$ ($\eta=10$, $100$).
Again, the pressure along the $x$-axis is compared with the asymptotic forms (\ref{eq: 3.2.3}), (\ref{eq: 3.2.1}).
Figure \ref{fig: 5} shows the same data  for the diagonally aligned square of figure \ref{fig: 1}({\bf d}), with $L=1$.
These pressure fields both have a component which is strongly concentrated close to the point of 
minimum separation, but $P_\theta$ is elevated across the entire disc and square.
In the case of the disc, there is fair agreement with the approximate theory of section \ref{sec: 3}, with the pressure 
being reduced due to fluid leaking from above and below the point of closest approach. In the case of the 
diagonally aligned square, the fluid has a shorter path to escape and the magnitude of the pressure is greatly
reduced. 

\section{Estimates of generalised forces}
\label{sec: 5}

Section \ref{sec: 3} obtained asymptotic approximations for the component of the pressure field 
which diverge as the smallest gap, $Z$, approaches zero. What can be said about the generalised 
forces, which are expressed as integrals over the pressure field, as the edge contact condition, $Z=0$, is 
approached? Do the forces also diverge as $Z\to 0$, or is the singularity in the pressure field integrable?

In subsection \ref{sec: 5.1} below it will be argued that the pressure-field integrals are convergent, so that the 
generalised forces approach a finite limit as $Z\to 0$. It follows that the contributions of the asymptotic pressure
field obtained in section \ref{sec: 3} make contributions which are of the same order of magnitude as the 
remainder term $\Delta p(x,y)$ in equation (\ref{eq: 3.2.3}). The generalised forces are, therefore, determined 
by integrating the numerically determined pressure fields, obtained using the method described in section 
\ref{sec: 4}. (The divergent pressure field may still be relevant to determining generalised forces if the radius 
of curvature of the contacting edge diverges, $R\to \infty$, but that will not be pursued here). 

The generalised forces do, however, diverge as $\theta\to 0$ with $Z=0$. Section \ref{sec: 5.2} will give 
estimates for the dependence of the coefficients of the resistance matrix ${\bf R}$ (defined in equation (\ref{eq: 2.8})) on the tilt 
angle $\theta$, in the case where $Z\ll L\theta$. In section \ref{sec: 7} we discuss a phase diagram
which describes the long-time evolution of the system, regardless of whether the gap $Z$ is extremely small. 
By way of preparation, subsection \ref{sec: 5.3} considers the form of the resistance matrix in cases 
where $\eta=L\theta/Z$ is not assumed to be large.

\subsection{Convergence of integrals determining resistance matrix}
\label{sec: 5.1}

In section \ref{sec: 3}, the divergent pressure was expressed 
as a linear combination of two fields, $Q_1(x,y)$, $Q_2(x,y)$, which diverge as $Z\to 0$. 
Here we consider whether there are corresponding singular contributions to the generalised forces $F_z$, $F_x$ and $G_y$,  
which arise from estimating the integrals in equations (\ref{eq: 2.7}). 

The horizontal force component $F_x$ due to the shear stress in equation (\ref{eq: 2.3}) involves a pressure gradient, which after 
integrating by parts and using $\partial h/\partial x=\theta$ is expressed in terms of the vertical force component $F_z$ as
\begin{eqnarray}
\label{eq: 5.1.1}
F_x&=&-\int {\rm d}y \int {\rm d}x\ \left[\frac{\mu \dot X}{h}+\frac{1}{2}h\frac{\partial p}{\partial x}\right]-\theta F_z
\nonumber \\
&=&-\mu \dot X \int {\rm d}y\int {\rm d}x\ \frac{1}{h}+\frac{1}{2}\int {\rm d}y\int {\rm d}x\ p\frac{\partial h}{\partial x}-\theta F_z
\nonumber \\
&=&-\mu \dot X \int {\rm d}y\int {\rm d}x\ \frac{1}{h}-\frac{1}{2}\theta F_z
\end{eqnarray}
where we have assumed that the external pressure $p_0$ is zero.
Consider the component of $F_x$ which is independent of the pressure, for the case of a slightly curved leading 
edge where $f(y)=y^2/2R$, $R$ large, as illustrated in figure \ref{fig: 1}({\bf b}).  If  $Z=0$, this is proportional to the integral
\begin{eqnarray}
\label{eq: 5.1.25}
\int {\rm d}y\int {\rm d}x\ \frac{1}{h}&=&\frac{1}{\theta}\int_{-W/2}^{W/2} {\rm d}y\int_0^L {\rm d}x\ \frac{1}{f(y)+x}
=\frac{1}{\theta}\int_{-W/2}^{W/2}{\rm d}y \left[\ln\left(1+\frac{2LR}{y^2}\right)\right]
\nonumber \\
&=&\frac{W}{\theta}\left(\ln\left(1+{\cal R} \right)+ 2\sqrt{{\cal R} }\arctan\left({\cal R} ^{-1/2}\right)\right)
\end{eqnarray}
where ${\cal R}=8RL/W^2$ is a dimensionless measure of the radius of curvature of the leading edge.
So the integral remains finite as $Z\to 0$, so long as the leading edge has some 
curvature, i.e. $R$ is finite.

Equation (\ref{eq: 5.1.25}) shows that one of the area integrals which determine the generalised 
forces remains finite as $Z\to 0$. This is the case in spite of the fact that the integral over $x$ has 
a weak, logarithmic, divergence along the $x$-axis. The same observation applies to integrals of the pressure fields, 
$Q_1(x,y)$ and $Q_2(x,y)$. It is sufficient to show that the integrals are finite for $Z=0$. Here we consider only the case of 
the area integral of $Q_1(x,y)$. A similar approach verifies that all of the other integrals remain finite at $Z=0$.
Considering the evaluation of the vertical force $F_z$ resulting from a pressure field $Q_1(x,y)$, with $Z=0$ and $f(y)=y^2/2R$, 
we have:
\begin{eqnarray}
\label{eq: 5.1.55}
F_z&\propto&\frac{1}{\theta}\int_{-W/2}^{W/2} \mathrm{d}y\int_0^L\mathrm{d}x\left[\frac{x}{(f(y)+x)^2} - 
\frac{x^2}{L(f(y)+x)^2}\right] 
\nonumber \\
&=&\frac{W}{\theta}\left(\left(1+\frac{2}{3{\cal R}}\right)\ln\left(1+{\cal R}\right) + 
\frac{2}{3}\sqrt{{\cal R}}\arctan\left({\cal R}^{-1/2}\right) - \frac{2}{3}\right)
\ .
\end{eqnarray}
We find again that the expression evaluated for $Z=0$ is finite, even though the first $x$ integral here is 
logarithmically divergent along the $x$-axis. 

More generally, and for any of the plate shapes, it is useful to define the following set of integrals:
\begin{eqnarray}
\label{eq: 5.2.10}
{\cal H}&\equiv&\int_{{\cal A}}{\rm d}\mbox{\boldmath$x$}\ \frac{1}{h}
\nonumber \\
{\cal I}_{X}\equiv \int_{{\cal A}}{\rm d}\mbox{\boldmath$x$}\ P_{X}
&\quad\quad\quad&
{\cal J}_{X}\equiv \int_{{\cal A}}{\rm d}\mbox{\boldmath$x$}\ xP_{X} 
\nonumber \\
{\cal I}_{Z}\equiv \int_{{\cal A}}{\rm d}\mbox{\boldmath$x$}\ P_{Z}
&\quad\quad\quad&
{\cal J}_{Z}\equiv \int_{{\cal A}}{\rm d}\mbox{\boldmath$x$}\ xP_{Z} 
\nonumber \\
{\cal I}_{\theta}\equiv \int_{{\cal A}}{\rm d}\mbox{\boldmath$x$}\ P_{\theta}
&\quad\quad\quad&
{\cal J}_{\theta}\equiv \int_{{\cal A}}{\rm d}\mbox{\boldmath$x$}\ xP_{\theta}
\end{eqnarray}
The argument leading to equation (\ref{eq: 5.1.25}) shows that ${\cal H}$ remains finite
as $Z\to 0$, and equation (\ref{eq: 5.1.55}) indicates that ${\cal I}_X$ and ${\cal I}_Z$ 
both remain finite in this limit. The same approach applied to integrals over the fields 
$xQ_1$, $Q_2$, and $xQ_2$ shows that these are also finite at $Z=0$, so that all of the integrals 
in (\ref{eq: 5.2.10}) remain finite in the limit as $Z\to 0$. The singular pressure field which arises in the 
$Z\to 0$ limit, does not, therefore, result in a singular contribution to any of the generalised forces. 
The generalised forces can, therefore, in this $Z\to 0$ limit, be determined numerically using the method of section \ref{sec: 4}.

We remark that ${\cal H}$ is clearly positive, while ${\cal I}_Z$, ${\cal I}_\theta$, ${\cal J}_Z$ and
${\cal J}_\theta$ are all negative because reducing either $Z$ or $\theta$ must cause a positive pressure to 
drive fluid out of the gap. Increasing $X$ must cause positive pressure, hence ${\cal I}_X$, ${\cal J}_X$ must be positive.

\subsection{Resistance matrix close to contact}
\label{sec: 5.2}

In section \ref{sec: 5.1} it was argued that, for $\theta\ll 1$, in the limit as $Z\to 0$, 
the force coefficients become independent of $Z$. 
Furthermore, examination of (\ref{eq: 5.1.25}) and (\ref{eq: 5.1.55}), and of similar calculations for the other integrals, 
shows that the dependences of these integrals on the tilt angle $\theta$ are 
\begin{equation}
\label{eq: 5.1.60}
\int_{{\cal A}}{\rm d}\mbox{\boldmath$x$}\ \frac{1}{h}\sim \frac{L}{\theta}
\ ,\ \ \ 
\int_{{\cal A}}{\rm d}\mbox{\boldmath$x$}\ Q_1(\mbox{\boldmath$x$})\sim \frac{L}{\theta}
\ ,\ \ \ 
\int_{{\cal A}}{\rm d}\mbox{\boldmath$x$}\ xQ_1(\mbox{\boldmath$x$})\sim \frac{L^2}{\theta}
\ ,\ \ \ 
\int_{{\cal A}}{\rm d}\mbox{\boldmath$x$}\ Q_2(\mbox{\boldmath$x$})\sim L^2
\ ,\ \ \ 
\int_{{\cal A}}{\rm d}\mbox{\boldmath$x$}\ xQ_2(\mbox{\boldmath$x$})\sim L^3
\end{equation}
where we've taken $L$ as the generic length scale.

Taking together the asymptotic result (\ref{eq: 3.2.3}), 
equation (\ref{eq: 5.1.1}), and the definitions (\ref{eq: 5.1.60}), we express equation (\ref{eq: 2.8}) in the form
\begin{eqnarray}
\label{eq: 5.2.1}
F_x&=&\mu L^2 \left[\frac{a_{11}}{\theta}\frac{\dot X}{L}+\frac{a_{12}}{\theta^2}\frac{\dot Z}{L}+\frac{a_{13}}{\theta^2}\dot\theta\right]
\nonumber \\
F_z&=&\mu L^2\left[\frac{a_{21}}{\theta^2}\frac{\dot X}{L}+\frac{a_{22}}{\theta^3}\frac{\dot Z}{L}+\frac{a_{23}}{\theta^3}\dot\theta\right]
\nonumber \\
G_y&=&\mu L^3 \left[\frac{a_{31}}{\theta^2}\frac{\dot X}{L}+\frac{a_{32}}{\theta^3}\frac{\dot Z}{L}+\frac{a_{33}}{\theta^3}\dot\theta\right]
\ .
\end{eqnarray}
where all of the $a_{ij}$ are dimensionless numbers, depending on the geometry of the settling surface. 
The elements $a_{i1}$ are determined by numerically computing the pressure field after 
setting $\dot{X}=1$, $\dot Z=\dot \theta=0$, and then performing the integrals in equation (\ref{eq: 2.7}), 
proceeding similarly for the $a_{i2}$ and $a_{i3}$. The arguments in section \ref{sec: 5.1}
indicate that the $a_{ij}$ approach a finite limit as $Z\to 0$.

As a consequence of the Lorentz reciprocal theorem (see \cite{Kim+91} chapter 2 section 2.3 and chapter 5 section 5.2) the 
resistance matrix relating the forces / torques on a rigid body in a Stokes flow to its translational / rotational velocities is 
symmetric, so the form of the equation (\ref{eq: 5.2.1}) shows that the $3\times 3$ matrix $\{a_{ij}\}$ is symmetric. 
Moreover, the form of the Reynolds equations implies various other relations between the elements $a_{ij}$.
Note that the inhomogeneous term in (\ref{eq: 4.2}) is the same for the responses to $\dot X$ and $\dot Z$ 
apart from a factor of $-\theta/2$, implying that there are two relations between the first and second columns of the matrix 
$\{a_{ij}\}$, reducing the number of independent elements to four. This matrix will be written in the form
\begin{equation}
\label{eq: 5.2.2}
{\bf a}=
\left(\begin{array}{ccc}
-a&b&d\cr
b&-2b&-2d\cr
d&-2d&-c
\end{array}\right)
\end{equation}
where the $\{a,b,c,d\}$ are positive numbers, as shown below.

The coefficients in (\ref{eq: 5.2.2}) can be expressed in terms of the integrals (\ref{eq: 5.2.10}):
\begin{equation}
\label{eq: 5.2.12}
a=\frac{\theta}{L}{\cal H}-\frac{\theta^3}{4\mu L}{\cal I}_{Z}
\ ,\ \ \ 
b=-\frac{\theta^3}{2\mu L}{\cal I}_{Z}
\ ,\ \ \ 
c=-\frac{\theta^3}{\mu L^3}{\cal J}_{\theta}
\ ,\ \ \ 
d=-\frac{\theta^3}{2\mu L^2}{\cal J}_{Z}=-\frac{\theta^3}{2\mu L^2}{\cal I}_\theta
\end{equation}
where the ${\cal I}_X=-\frac{\theta}{2}{\cal I}_Z$ was used to derive the first expression in (\ref{eq: 5.2.12}).
The relationship ${\cal J}_Z={\cal I}_\theta$ is a consequence of the Lorentz reciprocal theorem \cite{Kim+91}.

As well as equalities between elements of the resistance matrix, there are also various 
inequalities. Therefore, for our 
planforms ${\cal A}$ where $0<x<L$ (which includes all of the examples in figure \ref{fig: 1}), we have 
\begin{equation}
\label{eq: 5.2.11}
L|{\cal I}_{Z}|>|{\cal J}_{Z}|
\ ,\ \ \ 
L|{\cal I}_{\theta}|>|{\cal J}_{\theta}|
\ .
\end{equation}
Combining (\ref{eq: 5.2.12}) with (\ref{eq: 5.2.11}), we can derive three inequalities between the parameters $\{a,b,c,d\}$:
\begin{equation}
\label{eq: 5.2.13}
b<2a
\ ,\ \ \ 
c<2d
\ ,\ \ \ 
d<b
\end{equation}
which when combined with (\ref{eq: 5.2.12}) shows that all four parameters are positive, as claimed.

We can make further statements about the relative magnitudes of the numbers $\{a,b,c,d\}$ in the case where 
the left-hand edge is well-approximated by a straight line (for the object in figure \ref{fig: 1}({\bf b}), in the limit as $R\to \infty$). 
In this case, the pressure at the left-hand edge has a component, represented by the function $Q_1$ in equation (\ref{eq: 3.2.1}), 
having a maximum value which diverges as $Z\to 0$. The displacement of this maximum from the left-hand edge is $Z/\theta$, 
and the function has a $\sim 1/x$ decay on the right-hand side of its maximum. These considerations indicate that the integrals 
${\cal I}_X$ and ${\cal I}_Z$ have a logarithmic divergence as the radius of curvature at the contact point, $R$, approaches 
infinity. However, because of the factor of $x$ in the integrals defining ${\cal J}_X$, ${\cal J}_Z$ and ${\cal J}_\theta$, 
these latter integrals do not diverge as $R\to \infty$. Hence, we conclude that
\begin{equation}
\label{eq: 5.2.14}
\{a,b\}\gg \{c,d\}
\quad\quad\quad
{\bf radius}\ {\bf of}\ {\bf curvature}\ {\bf at}\ {\bf contact}\ {\bf point}\;R\to \infty
\end{equation}
i.e., the parameters $\{a,b\}$ will generally be much greater than the parameters $\{c,d\}$ when the radius
of curvature close to the minimal gap is large.

Using the numerical procedure described in section \ref{sec: 4}, and setting $\mu=1$, we computed the pressure fields and generalised forces for the 
rectangle with a curved edge, for the disc,  and for diagonal square with a minimal gap of $Z=10^{-7}$.
For the rectangle with a curved edge, figure \ref{fig: 1}({\bf b}) (with $L=1$ and $W=10$), we found
\begin{equation}
\label{eq: 5.2.10a}
{\bf a}\approx
\left(\begin{array}{ccc}
-105&129&21.7\cr
129&-259&-43.3\cr
21.7&-43.3&-13.2
\end{array}\right)
\ ,\ \ \ 
{\bf b}\approx
\left(\begin{array}{ccc}
-0.0247&-0.0124&0\cr
-0.0124&-0.0148&0.0281\cr
0&0.0281&-0.168
\end{array}\right)
\ ,\ \ \ ({\bf Rectangle}\ R=10)
\end{equation}
\begin{equation}
\label{eq: 5.2.10b}
{\bf a}\approx
\left(\begin{array}{ccc}
-191&255&27.5\cr
255&-510&-54.9\cr
27.5&-54.9&-14.2
\end{array}\right)
\ ,\ \ \ 
{\bf b}\approx
\left(\begin{array}{ccc}
-0.0158&-0.00790&0\cr
-0.00790&-0.00730&0.0129\cr
0&0.0129&-0.120
\end{array}\right)
\ ,\ \ \ ({\bf Rectangle}\ R=100)
\end{equation}
Note that in fact \{$a,b\}$ here are larger than $\{c,d\}$, by factors which increase with $R$.

For the disc (radius $R=0.5$), the approximate values of the matrix ${\bf a}=\{a_{ij}\}$ and its inverse ${\bf b}$ were found to be
\begin{equation}
\label{eq: 5.2.3}
{\bf a}\approx
\left(\begin{array}{ccc}
-6.56&6.93&0.994\cr
6.93&-13.86&-1.99\cr
0.994&-1.99&-0.570
\end{array}\right)
\ ,\ \ \ 
{\bf b}\approx
\left(\begin{array}{ccc}
-0.323&-0.162&0\cr
-0.162&-0.225&0.503\cr
0&0.503&-3.51
\end{array}\right)
\ ,\ \ \ ({\bf Disc})
\end{equation}
so that for the disk $a\approx 6.56$, $b\approx 6.93$, $c\approx 0.570$, $d\approx 0.994$. 
For the diagonal square (with $L=1$), we found
\begin{equation}
\label{eq: 5.2.4}
{\bf a}\approx
\left(\begin{array}{ccc}
-1.90&1.04&0.290\cr
1.04&-2.08&-0.580\cr
0.290&-0.580&-0.214
\end{array}\right)
\ ,\ \ \ 
{\bf b}\approx
\left(\begin{array}{ccc}
-0.723&-0.361&0\cr
-0.361&-2.12&5.26\cr
0&5.26&-18.9
\end{array}\right)
\ ,\ \ \ ({\bf Diagonal}\ {\bf square})
\end{equation}
so that in this case $a\approx 1.90$, $b\approx 1.04$, $c\approx 0.214$, $d\approx 0.290$.
Note that in all four cases the coefficients satisfy the inequalities contained in (\ref{eq: 5.2.13}). 
Using approximately $10^4$ elements in different Delaunay triangularisations, our results were 
consistent within a tolerance of better than $2\%$. The details of the implementation of the method, 
its generalisations, and an assessment of its accuracy will be discussed in later work.

\subsection{General form for the resistance matrix}
\label{sec: 5.3}

\begin{figure}[h]
\includegraphics[width=0.5\textwidth]{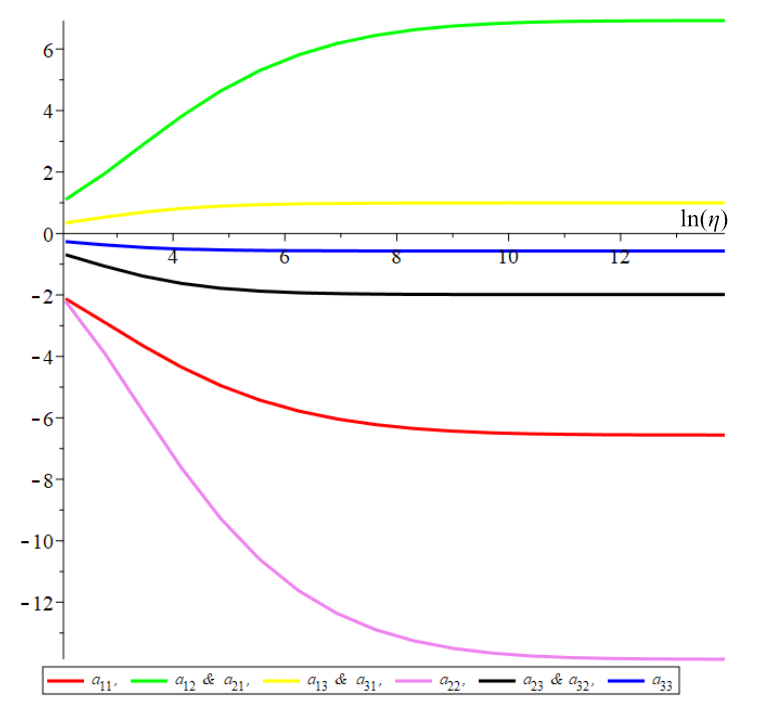}
\caption{Plots of $a_{ij}(\eta)$, as a function of $\ln\eta$.}
\label{fig: 6}
\end{figure}

As well as considering the case where the edge is close to contact, in section \ref{sec: 7} we shall also 
consider a phase diagram, which describes the long-time behaviour as a function of the initial coordinates of the 
system. Equation (\ref{eq: 5.2.1}) was obtained subject to the assumption that $Z\ll L\theta$, but in order to describe the 
motion for general initial conditions we require a form for the resistance matrix which does not depend
upon this assumption. In \cite{Wil+23}, it was shown that it useful to introduce a dimensionless variable 
\begin{equation}
\label{eq: 5.3.1}
\eta=\frac{L\theta}{Z}
\ ,
\end{equation}
which allows us to express the generalised forces in the form
\begin{eqnarray}
\label{eq: 5.3.2}
\frac{F_x}{\mu L^2}&=&\frac{L}{Z}\left[A_{11}(\eta)\frac{\dot X}{L}+A_{12}(\eta)\frac{\dot Z}{Z}+A_{13}(\eta)\dot \eta\right]
\nonumber \\
\frac{F_z}{\mu L^2}&=&\frac{L^2}{Z^2}\left[A_{21}(\eta)\frac{\dot X}{L}+A_{22}(\eta)\frac{\dot Z}{Z}+A_{23}(\eta)\dot\eta\right]
\nonumber \\
\frac{G_y}{\mu L^3}&=&\frac{L^2}{Z^2}\left[A_{31}(\eta)\frac{\dot X}{L}+A_{32}(\eta)\frac{\dot Z}{Z}+A_{33}(\eta)\dot\eta\right]
\ .
\end{eqnarray}
Comparison with (\ref{eq: 5.2.1}) shows that the coefficients $a_{ij}$ are also functions of $\eta$, with 
the following relations between the coefficients $A_{ij}$ and $a_{ij}$:
\begin{equation}
\label{eq: 5.3.3}
\left(\begin{array}{ccc}
a_{11}(\eta)&a_{12}(\eta)&a_{13}(\eta)\cr
a_{21}(\eta)&a_{22}(\eta)&a_{23}(\eta)\cr
a_{31}(\eta)&a_{32}(\eta)&a_{33}(\eta)
\end{array}\right)
= 
\left(\begin{array}{ccc}
\eta A_{11}(\eta)     & \eta^2[A_{12}(\eta)-\eta A_{13}(\eta)]  & \eta^2A_{13}(\eta)\cr
\eta^2A_{21}(\eta) & \eta^3[A_{22}(\eta)-\eta A_{23}(\eta)] & \eta^3A_{23}(\eta)\cr
\eta^2A_{31}(\eta) & \eta^3[A_{32}(\eta)-\eta A_{33}(\eta)] & \eta^3A_{33}(\eta)
\end{array}\right)
\ .
\end{equation}
Equations (\ref{eq: 5.2.1}) and (\ref{eq: 5.3.2}) are two alternative representations of 
(\ref{eq: 2.8}) which have different advantages. Equation (\ref{eq: 5.2.1}) has coefficients 
$a_{ij}(\eta)$ which are symmetric, and remain finite as $\eta\to \infty$. Equation (\ref{eq: 5.3.2})
will prove useful when we investigate the phase diagram in section \ref{sec: 7} by obtaining
an autonomous equation for $\dot \eta$.
In figure \ref{fig: 6} we plot the $a_{ij}(\eta)$ coefficients 
for the case of the disc with $R=1/2$, and note that the curves approach the values in (\ref{eq: 5.2.3}) as $\eta\to \infty$.
 
\section{Motion under gravity}  
\label{sec: 6}  
Consider sinking under gravity. If the plate makes no contact with the surface, the hydrodynamic forces and torque 
exactly balance the gravitational forces and torque, so then:
\begin{equation}
\label{eq: 6.1}
F_x=0
\,\ \ \ 
F_z={\cal W}
\,\ \ \ 
G_y=s{\cal W}L
\end{equation}
where ${\cal W}$ is the weight of the upper object.
After contact, the vertical force is
\begin{equation}
\label{eq: 6.2}
F_z={\cal W}-{\cal N}
\ ,\ \ \ 0\le {\cal N}\le {\cal W}
\end{equation}
where ${\cal N}$ is the reaction force at the point of contact. The torque is unchanged because
the contact point lies on the reference axis

After contact, when a frictional force arises, the horizontal force balance is:
\begin{equation}
\label{eq: 6.3}
F_x= {\rm sign}(\dot X)\;\nu_{\rm dyn}{\cal N}\quad ,
\quad\dot X\ne 0
\end{equation}
or, if sliding does not occur, the horizontal component of the hydrodynamic force must satisfy:  
\begin{equation}
\label{eq: 6.3a}
|F_x|\le \nu_{\rm stat}{\cal N}\quad ,\quad\dot X=0
\end{equation}
where $\nu_{\rm dyn}$ and $\nu_{\rm stat}$ are (respectively) dynamic and static coefficients of friction, 
with $\nu_{\rm stat}> \nu_{\rm dyn}>0$. If there is contact without sliding, then $\dot X=\dot Z=0$, 
and equation (\ref{eq: 5.2.1}) indicates that $F_x=-\theta F_z/2$, so that (\ref{eq: 6.3a}) is always satisfied 
as $\theta\to 0$.

Three scenarios will be considered in turn: motion prior to an edge making contact, motion after contact without 
slipping, and slipping contact. The discussion will make some assumptions about signs and other properties 
of matrix elements $a_{ij}$ and $b_{ij}$. The text will give a numbered list these assumptions as they arise.
Their validity will be addressed in subsection \ref{sec: 6.4}, which also summarises our conclusions.

\subsection{Motion without contact}
\label{sec: 6.1}

It will be convenient to define a timescale $T$:
\begin{equation}
\label{eq: 6.1.1}
T=\frac{\mu L^2}{{\cal W}}
\ .
\end{equation}
Inverting the linear system (\ref{eq: 5.2.1}) leads to a system of equations 
of the form
\begin{equation}
\label{eq: 6.1.2}
\left(\begin{array}{c}
\dot X/L\cr
\dot Z/L\cr
\dot \theta
\end{array}\right)
=\frac{1}{T}
\left(\begin{array}{ccc}
b_{11}\theta &b_{12}\theta^2 &0\cr
b_{21}\theta^2 &b_{22}\theta^3 &b_{23}\theta^3\cr
0&b_{32}\theta^3 &b_{33}\theta^3
\end{array}\right)
\left(\begin{array}{c}
0\cr
1\cr
s
\end{array}\right)
\end{equation}
where the dimensionless coefficients $b_{ij}$ are the matrix elements of the inverse matrix ${\bf b}={\bf a}^{-1}$. 
The structure of the matrix ${\bf a}$ implies that $b_{13}=b_{31}=0$ exactly, and also that ${\bf b}$ is symmetric, $b_{ji}=b_{ij}$.

Consider first the equation of motion for $\theta$, which is independent of the other variables:
\begin{equation}
\label{eq: 6.1.3}
\dot \theta=-\frac{C}{T}\theta^3
\end{equation}
where
\begin{equation}
\label{eq: 6.1.4}
C=-b_{32}-s b_{33}
\ .
\end{equation}
We assume that $b_{32}>0$ and that $b_{33}<0$ ({\bf assumption 1}) , so that 
$C<0$ for small $s$, becoming positive when $s$ exceeds some critical value, 
\begin{equation}
\label{eq: 6.1.45}
s_1\equiv -b_{32}/b_{33}
\ .
\end{equation}
If $C<0$, then $\theta(t)$ is increasing. However, the centre of mass must be sinking, that is 
\begin{equation}
\label{eq: 6.1.11a}
\dot Z+sL \dot \theta<0
\ .
\end{equation}
That requires 
\begin{equation}
\label{eq: 6.1.12}
b_{22}+2sb_{23}+s^2b_{33}<0
\end{equation}
for all $s\in [0,1]$. If $\theta$ is increasing because $C<0$, then left-hand edge
must be sinking. In this situation, contact must be made in a finite time.

The case where $C>0$ (defined by (\ref{eq: 6.1.4})) so that $\theta(t)$ is decreasing, is more complex. 
There are three possibilities:

\begin{enumerate}

\item The object may make contact with the surface in finite time $t^\ast$, so that $Z(t^\ast)=0$
while $\theta(t^\ast)>0$. The subsequent motion is then one or other of the cases considered in 
subsections \ref{sec: 6.2} or \ref{sec: 6.3} below.

\item The object may sink so that both $Z(t)$ and $\theta(t)$ decrease as $t\to \infty$, without contact  
ever happening.

\item The equations of motion (\ref{eq: 5.2.1}) were obtained under the assumption that $L\theta/Z\gg 1$.
If $\theta\to 0$ sufficiently rapidly, this assumption ceases to be valid, and equations (\ref{eq: 5.2.1}) may 
cease to be an appropriate tool for analysing the motion.

\end{enumerate}

Considering the case where $\theta(t)$ is decreasing in greater detail, the solution of (\ref{eq: 6.1.3}) is
\begin{equation}
\label{eq: 6.1.5}
\theta=\frac{\theta_0}{\sqrt{1+\frac{2C\theta_0^2t}{T}}}
\end{equation}
where $\theta_0=\theta(0)$. Thus $\theta(t)$ decreases, as $t^{-1/2}$ if $C>0$ when $t\to \infty$. 
Next consider the evolution of the gap, $Z(t)$, assuming $C>0$. Given the solution for $\theta(t)$, $Z(t)$ satisfies
\begin{equation}
\label{eq: 6.1.6}
\dot Z=\frac{(b_{22}+sb_{23})L}{T}\theta^3=B\left(1+\alpha t\right)^{-3/2}
\end{equation}
where
\begin{equation}
\label{eq: 6.1.7}
B=\frac{(b_{22}+sb_{23})L\theta_0^3}{T}
\ ,\ \ \ 
\alpha=\frac{2C\theta_0^2}{T}
\ .
\end{equation}
The solution is
\begin{eqnarray}
\label{eq: 6.1.8}
Z&=&Z_0+\frac{2B}{\alpha}\left[1-\frac{1}{\sqrt{1+\alpha t}}\right]
\nonumber \\
&=&Z_0-DL\theta_0\left[1-\frac{1}{\sqrt{1+\alpha t}}\right]
\end{eqnarray}
where
\begin{equation}
\label{eq: 6.1.9}
D=\frac{b_{22}+sb_{23}}{b_{32}+sb_{33}}
\end{equation}
As $t\to \infty$:
\begin{equation}
\label{eq: 6.1.10}
Z(t)\to Z_0-DL\theta_0
\ .
\end{equation}
If this quantity is negative, there is contact in finite time.
If this is positive, the prediction is that the object ceases to
sink, which is physically an absurdity. However, we should note that  the 
derivation of (\ref{eq: 5.2.1}) assumed that $Z_0$ is very small,
so that $Z_0\ll L\theta_0$. We conclude that if $D>0$ contact must occur in 
finite time. Because we already assumed that $C>0$, and $-C$ is the denominator  
of (\ref{eq: 6.1.9}), this condition for contact in finite time is $b_{22}+sb_{23}<0$.
Noting that as we have assumed $b_{32}>0$, (and $b_{23}=b_{32}$), then contact occurs in finite time if
\begin{equation}
\label{eq: 6.1.11}
s<s_3\equiv \frac{-b_{22}}{b_{23}}
\quad\quad\quad
{\bf contact}\ {\bf in}\ {\bf finite}\ {\bf time}
\ .
\end{equation}
If $s>s_3$, the value of $\eta\equiv L\theta/Z$ decreases, and equations 
(\ref{eq: 5.2.1}) cease to be applicable. The motion may then approach a settling with a constant
value of $\eta$, or the value of $\theta$ may become negative, with the opposite end of the object 
approaching or contacting the surface. These cases will be considered in section \ref{sec: 7}.

\subsection{Contact, no sliding}
\label{sec: 6.2}

Once the object makes contact with the lower surface the force equations are altered, but the torque 
about the point of contact is still given by (\ref{eq: 6.1}). If upon contact there is no sliding, then $\dot X=\dot Z=0$, 
and equation (\ref{eq: 5.2.1}) for the torque gives us
\begin{equation}
\label{eq: 6.2.1}
\frac{s}{T}=a_{33}\frac{\dot \theta}{\theta^3}
\ .
\end{equation}
The solution is
\begin{equation}
\label{eq: 6.2.2}
\theta(t)=\frac{\theta_0}{\sqrt{1-\frac{2s\theta_0^2t}{a_{33}T}}}
 .
\end{equation}
The equation for the vertical force is then
\begin{equation}
\label{eq: 6.2.3}
\frac{1}{T}\left[1-\frac{{\cal N}}{{\cal W}}\right]=a_{23}\frac{\dot \theta}{\theta^3}=\frac{a_{23}}{a_{33}}\frac{s}{T}
\ .
\end{equation}
This gives
\begin{equation}
\label{eq: 6.2.4}
\frac{{\cal N}}{{\cal W}}=1-\frac{a_{23}}{a_{33}}s
\ .
\end{equation}
Noting that, according to (\ref{eq: 5.2.2}), $a_{23}$ and $a_{33}$ have the same sign, the 
requirement that ${\cal N}>0$ leads to
\begin{equation}
\label{eq: 6.2.5}
s<s_2\equiv \frac{a_{33}}{a_{23}}
\quad\quad\quad
{\bf no}\ {\bf slip}\ {\bf contact}
\ .
\end{equation}

\subsection{Contact with sliding}
\label{sec: 6.3}

Consider equation (\ref{eq: 5.2.1}) with $Z=\dot{Z}=0$. We shall assume that $\dot X<0$ (that is, the upper
object slides to the left). For the contact with sliding case then 
$F_x=-\nu {\cal N}$ and $F_z={\cal W}-{\cal N}$ where ${\cal N}$ is the reaction force, and $\nu\equiv\nu_{\rm dyn}$ 
the coefficient of dynamic friction, 
so we have: 
\begin{eqnarray}
\label{eq: 6.3.1a}
-\nu {\cal N}&=&\mu L^2 \left[\frac{a_{11}}{\theta}\frac{\dot X}{L}+\frac{a_{13}}{\theta^2}\dot\theta \right]
\nonumber \\
{\cal W}-{\cal N}&=&\mu L^2 \left[\frac{a_{21}}{\theta^2}\frac{\dot X}{L}+\frac{a_{23}}{\theta^3}\dot\theta \right]
\nonumber \\
{\cal W}s&=&\mu L^2 \left[\frac{a_{31}}{\theta^2}\frac{\dot X}{L}+\frac{a_{33}}{\theta^3}\dot\theta \right]
\ .
\end{eqnarray}
We eliminate $\dot X$ and $\dot \theta$ from these equations, to yield an expression for ${\cal N}$.
We assume that $\nu $ is strictly positive, and simplify by taking the limit $\theta \to 0$ where 
this is convenient. (Setting $\nu=0$ in our results may not be equivalent to considering the true
frictionless limit, $\nu\to 0$). After some algebra, we find
\begin{equation}
\label{eq: 6.3.2a}
{\cal N}=\theta {\cal W}\left[\frac{\alpha-\beta s}{\alpha\theta-\gamma\nu}\right]
\end{equation}
where 
\begin{equation}
\label{eq: 6.3.2b}
\alpha=(a_{13}a_{31}-a_{11}a_{33})
\ ,\ \ \ 
\beta=(a_{13}a_{21}-a_{23}a_{11})
\ ,\ \ \ 
\gamma=(a_{23}a_{31}-a_{33}a_{21})
\ .
\end{equation}
Imposing the condition $\theta \ll 1$, we have
\begin{equation}
\label{eq: 6.3.3a}
{\cal N}=\frac{\theta {\cal W}}{\gamma\nu}\left(\beta s-\alpha\right)
\ .
\end{equation}
Contact requires a non-negative ${\cal N}$ (contact no longer possible if a negative reaction is required to maintain it). 
The condition ${\cal N}>0$ is satisfied when
\begin{equation}
\label{eq: 6.3.3b}
s<s_4\equiv \frac{\alpha}{\beta}
\quad\quad\quad 
{\bf positive}\ {\bf reaction}\ {\bf force}
\end{equation}
provided $\beta/\gamma<0$ ({\bf assumption 2}): this is required so that ${\cal N}>0$ when $s<s_4$.

As well as requiring that the reaction force satisfies ${\cal W}>{\cal N}>0$, we should also
check that $\dot X<0$ (which was assumed in setting up equations (\ref{eq: 6.3.1a})), 
and that $\dot \theta<0$ (so that the object is sinking and energy is being dissipated). 
Note that, according to (\ref{eq: 6.3.2a}), the reaction force is proportional to 
$\theta$, and therefore becomes negligible as $\theta\to 0$. The final two equations 
of (\ref{eq: 6.3.1a}) therefore simplify to
\begin{eqnarray}
\label{eq: 6.3.1}
\frac{1}{T}&=&\frac{a_{21}}{\theta^2}\frac{\dot X}{L}+\frac{a_{23}}{\theta^3}\dot\theta
\nonumber \\
\frac{s}{T}&=&\frac{a_{31}}{\theta^2}\frac{\dot X}{L}+\frac{a_{33}}{\theta^3}\dot\theta
\ .
\end{eqnarray}
Solving these linear equations we obtain
\begin{equation}
\label{eq: 6.3.2}
\frac{\dot X T}{L\theta^2}=\frac{sa_{23}-a_{33}}{\gamma}
\ ,\ \ \ 
\frac{\dot \theta T}{\theta^3}=\frac{a_{31}-sa_{21}}{\gamma}
\ .
\end{equation}
We find that $\dot X<0$ when
\begin{equation}
\label{eq: 6.3.3}
s>s_5=\frac{a_{33}}{a_{23}} 
\quad\quad\quad
{\bf sliding}\ {\bf to}\ {\bf left}
\end{equation}
provided $a_{23}/\gamma<0$ ({\bf assumption 3}), so that $\dot X<0$ when $s<s_5$. Because 
$a_{23}<0$, if this condition is satisfied we must have $\gamma>0$.
Examination of the expression for $\dot \theta$ in (\ref{eq: 6.3.2}) then shows that 
this quantity is negative in the interval $s_5<s<s_4$, as required.  

\subsection{Summary and examples}
\label{sec: 6.4}

The discussion above shows that there are a variety of different behaviours 
when an edge is close to contact. It is valuable to collect the conclusions, and 
to supply numerical values for the boundaries between different regimes, in 
different geometries. We introduced a variety of different critical values
of the coordinate $s$ of the centre of mass, $s_1\ldots s_5$, which separate 
different dynamical behaviours. We will demonstrate that some of these 
are equal, and that in fact only $s_1$, $s_2$ and $s_3$ are distinct.

In the cases addressed in section \ref{sec: 6.1}, where the edge is not in contact,
the dynamics was specified in terms of the compliance matrix ${\bf b}={\bf a}^{-1}$.
It will be helpful to express the inequalities in terms of the positive numbers $\{a,b,c,d\}$ 
which are used to characterise the resistance matrix ${\bf a}$ in equation (\ref{eq: 5.2.2}).
For the inverse we have
\begin{equation}
\label{eq: 6.4.1}
{\bf b}=\frac{1}{(b-2a)(bc-2d^2)}\left(\begin{array}{ccc}
2bc-4d^2 & bc-2d^2 & 0 \cr
bc-2d^2   & ac-d^2 & d(b-2a) \cr
0 & d(b-2a) & b(2a-b)
\end{array}\right)
\ .
\end{equation}
Consider the value of $s_4$, defined by (\ref{eq: 6.3.3b}).
In terms of the $\{a, b, c, d\}$ constants from equation (\ref{eq: 5.2.2})
\begin{equation}
\label{eq: 6.3.5a}
s_4=\frac{\alpha}{\beta} =\frac{d^2-ac}{d(b-2a)}
\ .
\end{equation}
Compare this with $s_3$, defined in (\ref{eq: 6.1.11}). Using (\ref{eq: 6.4.1}) to express
this in terms of the coefficients in (\ref{eq: 5.2.2}), we have
\begin{equation}
\label{eq: 6.4.2} 
s_3=\frac{-b_{22}}{b_{23}}=\frac{d^2-ac}{d(b-2a)}=s_4
\ .
\end{equation}
Also, comparison of (\ref{eq: 6.2.5}) and (\ref{eq: 6.3.3}) shows that
\begin{equation}
\label{eq: 6.4.5}
s_5=s_2=\frac{a_{33}}{a_{23}}=\frac{c}{2d}
\end{equation}
and we remark that
\begin{equation}
\label{eq: 6.4.6}
s_1=-\frac{b_{32}}{b_{33}}=\frac{d}{b}
\ .
\end{equation}

For the circle and the diagonal square, the distinct critical values are
\begin{eqnarray}
\label{eq: 6.4.8}
s_1\approx 0.14<s_2\approx 0.29 <s_3\approx 0.45&\quad \quad & {\bf disc}
\nonumber \\
s_1\approx 0.28<s_2\approx 0.37 <s_3\approx 0.40&\quad \quad & {\bf diagonal}\ {\bf square}
\end{eqnarray}

Collating these results we conclude that, if there is initially no contact, then for $s<s_1$ the angle $\theta$ 
is increasing before contact occurs in a finite time. If $s_1<s<s_3$, then contact occurs in finite time,
with $\theta$ decreasing. If $s>s_3$, the left-hand edge does not contact in finite time: there may be no 
contact in finite time, or the right-hand edge may approach and possibly make contact.

\begin{figure}[h!]
\includegraphics[width=1.0\textwidth]{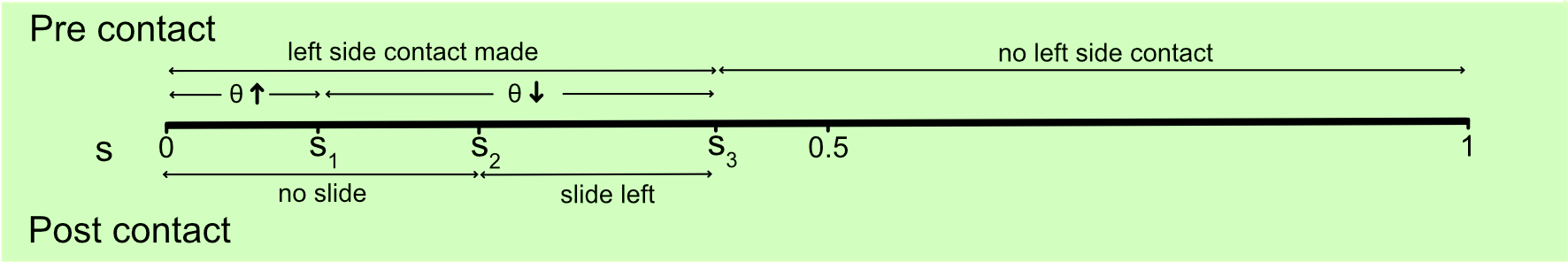}
\caption{Schematic illustration of the four possible behaviours of the edge-contact problem,
separated by three critical values of the centre of mass parameter $s\in[0,1]$. Above
the $s$-axis considers the initially non-contacting case: for $s<s_1$, $\theta $ increases until contact occurs in finite time, 
for $s_1<s<s_2$ contact in finite time with $\theta$ decreasing, for $s>s_3$: left edge does not make contact in finite time.
Below the $s$-axis considers motion after contact: for $s<s_2$, non sliding contact, for 
$s_2<s<s_3$, sliding contact.}
\label{fig: 7}
\end{figure}

If $s<s_3$, and left-hand contact does occur in finite time, we must consider what happens after contact.
We have shown, in section \ref{sec: 6.2}, that there is a non-sliding solution when $s<s_2$. In section 
\ref{sec: 6.3} we considered the possibility of a solution involving leftward sliding with contact, and found 
that this is possible when $s_2<s<s_3$. This solution requires that the reaction force ${\cal N}$ is positive, and 
that the sliding velocity $\dot X$ is negative. At $s=s_3$, the condition that ${\cal N}$ is positive ceases to be valid.
As $s$ approaches $s_2$ from above, the velocity $\dot X$ approaches zero, so there is no discontinuity 
with the non-sliding contact solution, which is valid for $s<s_2$. Our conclusions are illustrated in figure \ref{fig: 7}.

Note that our conclusions about the case where there is contact use only the assumption that the behaviour
of contacting surfaces is determined by conventional friction coefficients, $\nu_{\rm dyn}$ and $\nu_{\rm stat}$. 
This contrasts with the two-dimensional case considered in \cite{Wil+23}, where contact occurs along a line, rather
than at a point. In the case where there is contact along a line, we found that the solution must depend
upon assumptions about the microscopic structure of the contacting surfaces.

We should finally consider whether and how the various assumptions which were introduced in sections
\ref{sec: 6.1}-\ref{sec: 6.3} can be justified. Considering the assumptions in turn:

\begin{enumerate}

\item {\bf Assumption 1}: $b_{32}>0$ and $b_{33}<0$. From (\ref{eq: 6.4.1}), $b_{32}=d/(bc-2d^2)$. 
In cases where the radius of curvature at the point of contact is large, equation (\ref{eq: 5.2.14}) implies that $bc-2d^2>0$, 
so that $b_{32}>0$. Similarly, $b_{33}=-b/(bc-2d^2)<0$.

\item {\bf Assumption 2}: $\beta/\gamma<0$. First, $\beta=d(b-2a)<0$ by (\ref{eq: 5.2.13}). Next, when the radius of curvature at 
contact is large, $\gamma=bc-2d^2>0$, so that $\beta/\gamma <0$. In fact, we find that this inequality holds for all
of our examples, including the diagonal square, where $R=0$.

\item {\bf Assumption 3}: $a_{23}/\gamma<0$. Again, as $R\to \infty$, equation (\ref{eq: 5.2.14}) implies that $\gamma>0$, so that
$a_{23}/\gamma=-2d/\gamma<0$.

\end{enumerate}

In fact these assumptions are valid in all of the cases that we examined, including the diagonal
square (figure \ref{fig: 1}({\bf d})), where the radius of curvature is zero.

\section{Phase diagrams}
\label{sec: 7} 

The behaviour of the upper object, sinking under gravity, depends upon the position of the centre of gravity,
which is at a displacement $sL$ from the left-hand edge.  In section \ref{sec: 6}, we argued that, if the left 
hand edge is very close to contact, the edge makes contact in finite time if $s<s_3$. It is of interest to understand 
what happens under more general initial conditions. Addressing this this question is simplified by an observation 
which was made in \cite{Wil+23}, based upon the structure of equation (\ref{eq: 5.3.2}) above. Note that, in this representation, 
the elements of the resistance matrix are functions of a single coordinate, $\eta\equiv L\theta/Z$. We can invert 
this expression for the resistance matrix, writing
\begin{equation}
\label{eq: 7.1}
\frac{\mu L^4}{{\cal W}Z^2}
\begin{pmatrix}
\frac{\dot X}{L} \\ \frac{\dot Z}{Z} \\ \dot \eta\
\end{pmatrix}=
\begin{pmatrix}
B_{11}(\eta) & B_{12}(\eta) & B_{13}(\eta) \\
B_{21}(\eta) & B_{22}(\eta) & B_{23}(\eta) \\
B_{31}(\eta) & B_{32}(\eta) & B_{33}(\eta) 
\end{pmatrix}
\begin{pmatrix}
0 \\ 1 \\ s \cr
\end{pmatrix}
\end{equation}
where the $B_{ij}$ are elements of the inverse of $\{A_{ij}\}$. From this equation we see that 
$\eta(t)$ obeys an autonomous one-dimensional equation
\begin{equation}
\label{eq: 7.2}
\frac{\mu L^4}{{\cal W}Z^2}\frac{{\rm d}\eta}{{\rm d}t}=B_{32}(\eta)+sB_{33}(\eta)\equiv F_\eta(\eta)
\ .
\end{equation}
Having determined $\eta(t)$, the time dependences of other variables can be determined by integration 
(see discussion in \cite{Wil+23}).
The long-time dynamics of this system can be inferred from its set of fixed points, $\eta^\ast$, 
where 
\begin{equation}
\label{eq: 7.3}
F_\eta(\eta^\ast)=0
\ , \ \ \  
\left\{
\begin{array}{cc}
{\bf stable}: & F'_\eta(\eta^\ast)<0 \\
{\bf unstable}: & F'_\eta(\eta^\ast)>0
\end{array}
\right .
\ .
\end{equation}
We can display a phase diagram for the 
behaviour of the system, by plotting the locus of the fixed points 
in the $(s,\eta)$ plane. For a given system, the trajectories follow a line of constant $s$ towards 
a stable fixed point. Because the parameter $\eta$ approaches infinity
as $Z\to 0$, it is convenient to use an alternative variable $\phi$, defined by 
\begin{equation}
\label{eq: 7.4}
\phi=\frac{\eta}{2+\eta}
\ .
\end{equation}
This gives a much clearer diagrammatic representation: left-hand contact occurs for $\phi=+1$ 
and right-hand contact at $\phi=-1$. Fixed points with intermediate values of $\phi$ correspond to the object 
sinking with a constant value of the aspect ratio $\eta=L\theta/Z$. 
We expect that for, $s<s_3$, the line $\phi=+1$ is an attractive fixed point of $\phi$, and for 
$s>s_3$ it is a repeller. There is a line of fixed points in the $(s,\phi)$ plane emerging from 
$s=s_3$.

\begin{figure}[ht]
\includegraphics[width=0.8\textwidth]{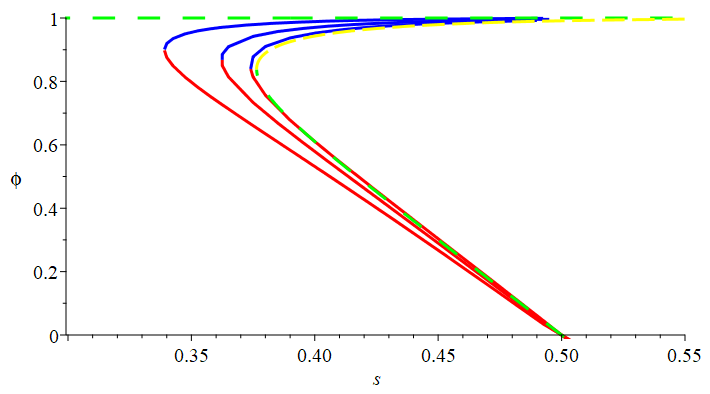}
\caption{Phase diagram showing fixed point $\phi^\ast$ for the finite rectangle of figure \ref{fig: 1}({\bf a}) 
as a function of the position of the centre of mass, $s$, for three different values of the aspect ratio (1, 2, 10, getting 
progressively closer to the infinite aspect curve), and for the infinite aspect 
ratio case obtained in \cite{Wil+23}. Solid red and blue represent stable and unstable fixed points for the 
finite rectangles respectively, dashed green and yellow for the infinite case}.
\label{fig: 8}
\end{figure}

We adopt the following approach to 
determining the phase diagram numerically. For every choice of $\phi^\ast$ (or $\eta^\ast$) except $\phi=1$, we obtain the single 
value of $s$ at which the line $\eta=\eta^\ast$ crosses the line of fixed points as follows. At our chosen value of $\eta$, we determine 
the coefficients $a_{ij}(\eta^\ast)$ in (\ref{eq: 5.2.1}) numerically.  Having obtained the $a_{ij}(\eta^\ast)$, 
we transform these to $A_{ij}(\eta)$ values via equation (\ref{eq: 5.3.3}), 
inverting this to give matrix $\mathbf{B}$ then finally using equation (\ref{eq: 7.2}) with 
$F_\eta(\eta)$ set to zero to give $s=-B_{32}/B_{33}$. By repeating this procedure for sufficient values of $\phi^\ast$ in 
the interval $[0,1)$ we were able to construct the phase diagrams presented below.  
(For the special case, the line $\phi=1$ is a fixed point for all $s$, stable for $s<s_3$.)

In \cite{Wil+23}, we were able to obtain analytical expressions for the matrix elements $A_{ij}$
and $B_{ij}$ for the case of a rectangle with aspect ratio $W/L\to \infty$.
As a control, figure \ref{fig: 8} shows the phase diagram 
for a rectangle with $W/L=1,2,10$. The phase line approaches the curve obtained in \cite{Wil+23} as
$W/L$ increases.

\begin{figure}[ht]
\includegraphics[width=0.8\textwidth]{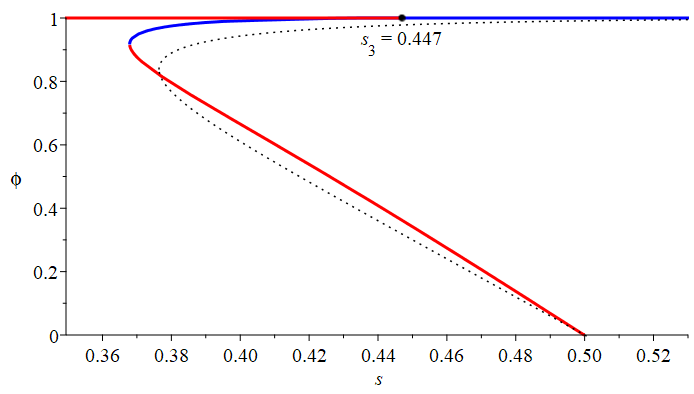}
\includegraphics[width=0.8\textwidth]{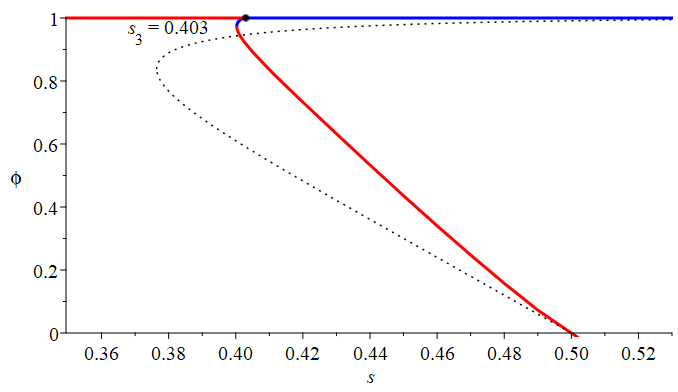}
\caption{Phase diagrams showing (above) fixed points $\phi^\ast$ for the disc 
of figure \ref{fig: 1}({\bf c}), and (below) for the diagonal square of 
figure \ref{fig: 1}({\bf d}). Solid red and blue represent stable and unstable fixed points. The infinite aspect 
ratio straight edge case obtained in \cite{Wil+23} is included for comparison (dotted black). For the diagonal square planform, there is a short 
region close to $s=s_3$ where the non-contacting fixed point line is unstable.}
\label{fig: 9}
\end{figure}

Figure \ref{fig: 9} shows the phase diagram for the disc (upper panel) and the diagonal square (lower). 
In each case, we include the phase diagram determined for the 
two-dimensional case treated in \cite{Wil+23} (equivalent to figure \ref{fig: 1}({\bf a}), with an infinite aspect ratio).

While being broadly similar, the phase diagrams for the examples treated in this paper do differ, in that the 
system treated in \cite{Wil+23} has an attractive fixed point at $\phi=+1$, for all $s$ (although the basin of attraction
may be exquisitely narrow), whereas the disc and diagonal square both have a region, $s>s_3$, where the 
left-hand side contact line $\phi=1$ is an unstable fixed point. 
There is a heuristic explanation for the difference in the form of the phase diagrams. 
In cases where the edge is curved, so that the gap can only close at one point on the edge. We can define a maximum 
value of $\eta$ for each point along the edge, $\eta_{\rm max}=L/f(y)$. It can then be argued that the phase
diagram will resemble that for an infinite straight edge, cut off at a value of $\eta$ which represents an appropriate average 
of $L/f(y)$. Equivalently, the phase diagram resembles that of the infinite straight edge, cut off at a value of $\phi$ 
which is slightly less than one.

\section{Concluding remarks}
\label{sec: 8}

This paper has investigated the implications of lubrication theory for fluid squeezed 
out of the gap between a pair of flat surfaces, in the case where there is very close 
approach at a point on the edge. To advance this study, we introduced a new efficient 
numerical method for solving the Reynolds equation (section \ref{sec: 4}), and asymptotic approximations
 to the pressure field close to the point of contact (section \ref{sec: 3}). We argued (section \ref{sec: 5}), that despite the 
 pressure fields diverging as contact is approached, the generalised forces do not become singular as the gap 
 closes.
 
 With these elements in place, we were able to investigate (section \ref{sec: 6}) the lubrication dynamics when an edge is close to 
 contact. This proved to be surprisingly complicated, as summarised in figure \ref{fig: 7}.
It is possible that the gap does not close in finite time, with the thickness of the fluid 
flow decreasing as $t^{-1/2}$. It is also possible for the gap to increase, with the surfaces eventually
contacting at another point.

The other possibility is that the gap closes at a point on the boundary. In this case there are two possibilities.
One is that the surfaces stick at their initial point of contact, and that fluid is extruded from the gap by the 
upper surface pivoting about the point of contact. In some cases, however, this solution is not viable. 
Satisfactory solutions can be found by allowing one surface to slide across the other, with a very weak reaction force. 
The sliding solution is only valid if the object is sliding to the left, so that frictional resistance acts to the right. 
The velocity of the sliding solution approaches zero at the point where the non-sliding solution 
becomes viable, so that there is a continuous transition between sliding and non-sliding contact 
as the position of the centre of mass is varied.

Sliding solutions were also discovered in the two-dimensional case analysed in \cite{Wil+23}, but there 
is a significant difference between the two cases. In the two-dimensional case, where contact occurs along a line
rather than at a point, it was found that the sliding solution depends upon the microscopic properties of the surface.
In the three-dimensional problem which we have addressed here, we find that the sliding solution is completely 
independent of the characteristics of the surfaces. 

A final comment: this work was motivated by the desire to understand what happens when a symmetric disc (a \lq coin') 
settles onto the base of a container full of viscous fluid. For the coin, we have $s=1/2>s_3$, so that the coin settles without contact 
in finite time. In fact, the phase diagram of figure \ref{fig: 9} shows that the coin will approach a flat configuration, $\theta=0$. 
If the coin is introduced at an angle which is not small, it may initially contact the base, but it will eventually lift off the surface.

\end{document}